\documentclass[journal]{IEEEtran}

\usepackage{cite}
\usepackage{graphicx}
\usepackage[font=footnotesize]{caption}
\usepackage[font=footnotesize]{subcaption}
\usepackage[latin1]{inputenc}
\usepackage[T1]{fontenc}
\usepackage{amsmath,amsfonts,amsbsy,amssymb}
\usepackage{mathabx} 
\usepackage{amssymb} 
\usepackage{amsmath}
\usepackage{amsthm}	
\usepackage{mathrsfs}
\usepackage[nolist]{acronym}
\usepackage{tabularx}
\usepackage{makecell}
\usepackage{multirow}
\usepackage{wasysym}
\usepackage{cancel}
\usepackage{float}
\usepackage{color}
\usepackage[lined, boxed, linesnumbered, ruled]{algorithm2e}
\usepackage{tikz}
\usetikzlibrary{arrows,calc,positioning}
\usetikzlibrary{arrows,trees}\usetikzlibrary{automata,positioning}
\usetikzlibrary{matrix,shapes,arrows,positioning,chains}
\usepackage{lettrine}
\usepackage{lipsum}

\usepackage{lipsum}
\usepackage{titlesec}

\titlespacing\section{0pt}{10.5pt plus 4pt minus 2pt}{3pt plus 2pt minus 2pt}
\titlespacing\subsection{0pt}{10.5pt plus 4pt minus 2pt}{3pt plus 2pt minus 2pt}
\titlespacing\subsubsection{0pt}{10.5pt plus 4pt minus 2pt}{3pt plus 2pt minus 2pt}

\usepackage{enumitem}
\begin{document}

\title{A Learning-Based Fast Uplink Grant for Massive IoT via Support Vector Machines and Long Short-Term Memory}

\author{
	\IEEEauthorblockN{Eslam Eldeeb, Mohammad Shehab, and Hirley Alves}
	\thanks{The authors are with Centre for Wireless Communications (CWC), University of Oulu, Finland. Email: firstname.lastname@oulu.fi.} 
	\thanks{This work is partially supported by Academy of Finland 6Genesis Flagship (Grant no. 318927), Aka Project EE-IoT (Grant no. 319008), FIREMAN (Grant no. 326301).} }
%
\maketitle

\begin{abstract}
The current random access (RA) allocation techniques suffer from congestion and high signaling overhead while serving massive machine type communication (mMTC) applications. To this end, 3GPP introduced the need to use fast uplink grant (FUG) allocation in order to reduce latency and increase reliability for smart internet-of-things (IoT) applications with strict QoS constraints. We propose a novel FUG allocation based on support vector machine (SVM), First, MTC devices are prioritized using SVM classifier. Second, LSTM architecture is used for traffic prediction and correction techniques to overcome prediction errors. Both results are used to achieve an efficient resource scheduler in terms of the average latency and total throughput. A Coupled Markov Modulated Poisson Process (CMMPP) traffic model with mixed alarm and regular traffic is applied to compare the proposed FUG allocation to other existing allocation techniques. In addition, an extended traffic model based CMMPP is used to evaluate the proposed algorithm in a more dense network. We test the proposed scheme using real-time measurement data collected from the Numenta Anomaly Benchmark (NAB) database. Our simulation results show the proposed model outperforms the existing RA allocation schemes by achieving the highest throughput and the lowest access delay of the order of 1 ms by achieving prediction accuracy of 98 $\%$ when serving the target massive and critical MTC applications with a limited number of resources.

\end{abstract}

\begin{IEEEkeywords}
	Alarm traffic, fast uplink grant, internet-of-things, machine learning, machine type communications, resource allocation, support vector machines.
\end{IEEEkeywords}
\vspace{-0mm}
\section{Introduction}\label{introduction}
\lettrine{C}{ELLULAR} communications have experienced a paradigm shift in recent years by introducing service modes dedicated to machine type communication (MTC), namely 
massive MTC (mMTC), and ultra-reliable low latency communications (URLLC) \cite{mahmood2019key}. Many smart IoT applications such as traffic control, autonomous vehicles, environmental monitoring, surveillance and crowd sensing are enabled by MTC \cite{IoT1,IoT2}. Due to the diversity among MTC scenarios, the quality-of-service (QoS) requirements vary. Therefore understanding their heterogeneous traffic behaviour becomes an essential part of any communication system \cite{6629817}. In this context, traffic modeling aims to capture the behaviour of the traffic using a probabilistic model that can be implemented easily and provide feasible means for efficient allocation of network resources. 

According to 
\cite{6629847}, MTC traffic, which is representative of many IoT applications, is classified into 3 different classes: \textit{a) Periodic Update (PU)}, which describes periodic traffic characterized by a small number of short packets, \textit{b) Event-Driven (ED)}, which describes non-periodic traffic due to a certain random trigger at an unknown time, and \textit{c) Payload Exchange (PE)}, which describes bursty traffic that usually comes after PU or ED traffic. Traffic models are classified into source traffic models and aggregated traffic models \cite{7389044}. Source traffic models, which treat every machine type communications device (MTD) as a single separate entity, are very accurate, though become extremely complex when modeling highly dense networks. Aggregated traffic models treat all MTDs within the network as one entity by simply accumulating all the traffic as one stream. When compared to the source traffic approach, the aggregated traffic models are less complex at cost of lower accuracy.
Authors in \cite{6629817} designed Coupled Markov Modulated Poisson Process (CMMPP) such that it captures the traffic behaviour efficiently based on a master node, called a background process, that describes the event. Authors in \cite{7996498} show that modeling CMMPP using multiple background processes is computationally expensive in time. They introduce Coupled Markovian Arrival Process (CMAP) model based on unicast and multicast distributions describing the regular traffic and traffic affected by events, respectively.


In the long-term evolution (LTE) systems, devices attempt to access the medium via random access (RA) procedures. Thus to obtain a radio resource and transmit a packet, user equipment (UE) undergoes a four-handshake procedure. This procedure suffers from high signaling overhead, which causes longer delays that prevent achieving the URLLC QoS requirements \cite{8705373}. It also fails to meet one of the most challenging requirements of smart IoT, namely is real-time performance due to high latency \cite{8401919}. Additionally, highly dense MTC deployments with hundreds of devices competing for meager resources will suffer from a large number of preamble collisions, which cause long delays and even packets drop \cite{FU}. To handle such issues, many solutions have been implemented as complementary features to the existing RA resource allocation algorithm, but they fail to guarantee MTC demands. Other researchers proposed credible techniques such as grant-free (GF) transmission, where each device chooses randomly a resource to transmit its packets without requests \cite{8877253}. Although GF solutions overcome the exchange of messages that causes signaling overhead problems, highly dense mMTC scenarios, where the number of devices is usually larger than the number of available resources, still suffer from a large number of collisions and resulting in longer delays. 

Recently, learning-based solutions have gained more attention to solve RA existing problems. In this context, fast uplink grant (FUG) was first introduced with the 2015 3rd generation partnership project (3GPP) technical report\cite{3gpp_fug}, and further discussed in \cite{FU}. It is a learning-based resource allocation technique, where the base station (BS) preemptively allocates resources to devices on a predictive basis. FUG reduces signaling overhead and completely removes collision. Introducing FUG in smart IoT applications allows for installing a large number of IoT devices within a limited number of frequency resources while satisfying their demands owing to efficient predictive algorithms \cite{8272467, 8401919}. Moreover, as FUG reduces signaling overhead and eliminates collisions, it economizes the energy consumption of IoT devices.

\section{Related Works and Contributions}\label{Related_Works}
In the following paragraphs, we discuss the recent works in the literature that address the resource allocation problem in IoT networks.

To address the high number of collisions in the GF allocation, authors in \cite{7823342} suggest a dynamic resource allocation scheme, that resolves the preamble collisions rather than avoiding them. Authors in \cite{article_NOMAGF} present a GF solution based on non-orthogonal multiple access (NOMA). Unfortunately, the proposed solutions suffer from undesired signaling overhead and collisions \cite{8712527}. Authors in \cite{shehab2020traffic} present a traffic prediction framework for IoT devices, which are influenced by binary Markovian events. A distributed non-orthogonal multiple access solution based on reinforcement learning is used in \cite{9119119}, while authors in \cite{9075198} present a sleeping multi-armed bandits (MAB) FUG solution. However, they focus only on achieving optimal resource allocation based on the QoS requirements of each device by employing an existing source traffic predictor and capturing the traffic behavior efficiently.

Authors in \cite{REV_BOOK} discussed the use of reinforcement learning and Q-tables to perform resource allocation. Mohammadi et al. \cite{8403658} proposed a multi-agent deep reinforcement learning technique, which performs distributed joint multi-resource
allocation. The authors applied transfer learning to the proposed model to speed up the convergence compared to applying deep Q-networks without transfer learning. Despite the advantages of applying reinforcement learning algorithms in resource allocation problems, it has two major disadvantages: \textit{1)} RL algorithms don't use any prior data about the network and only relies on reward signals, and \textit{2)} the convergence can be time-consuming.


Our contribution relies on the CMMPP traffic model as in \cite{6629817} to evaluate the performance of our proposed model compared to RA procedures. Different from \cite{7996498}, we introduce a model that handles multiple background processes, called the M-background processes CMMPP (M-CMMPP) model. This allows us to address congested network scenarios with a large number of active devices and a low number of resources without being computationally expensive. 
Furthermore, we propose a novel FUG algorithm based on 3 steps: 
\begin{enumerate}
    \item The first step is the device classification, whereby the BS prioritizes the devices according to their QoS requirements. To classify the devices, we employ a large-margin classifier based on support vector machines (SVMs), which search for the optimal decision boundary that is maximally far away from the closest points of each class in the training set \cite{10.5555/1394399}.
    \item The second step is the traffic prediction, which is typically a time-series problem. The BS monitors the activity behaviour of each MTD and forecasts which devices are active or inactive in a given transmission instant. In this context, long short-term memory (LSTM) is a special kind of recurrent neural networks (RNNs), that comprises 4 artificial neural networks (ANNs) layers, instead of just one, and acts as gates along with some point-wise operations \cite{article_LSTM}. These gates are trained to learn what to use from short and long past and what to forget. We apply LSTM for traffic prediction in our proposed FUG model. Moreover, we discuss the rationale for using LSTM and how it can be applied in real-time. 
    \item The last step is to exploit device classification and traffic prediction results to grant resources to devices that are worth. CMMPP and M-CMMPP models are introduced in the problem formulation, where we apply the proposed FUG algorithm and compare it to existing resource allocation techniques. Testing and simulation are performed using real-time measurement data collected from the Numenta Anomaly Benchmark (NAB) database.
\end{enumerate}

The rest of the paper is organized as follows. Section~\ref{system_model} presents the system model and problem formulation. In Section~\ref{FUG_alg_section}, we introduce the details of the proposed FUG solution. In Section~\ref{results_disc}, we present the simulation results and discussion. Finally, conclusion is presented in Section~\ref{conc_sec}.


\section{preliminaries}\label{system_model}

\begin{figure}[!t]
\centering
\includegraphics[width=0.96\columnwidth]{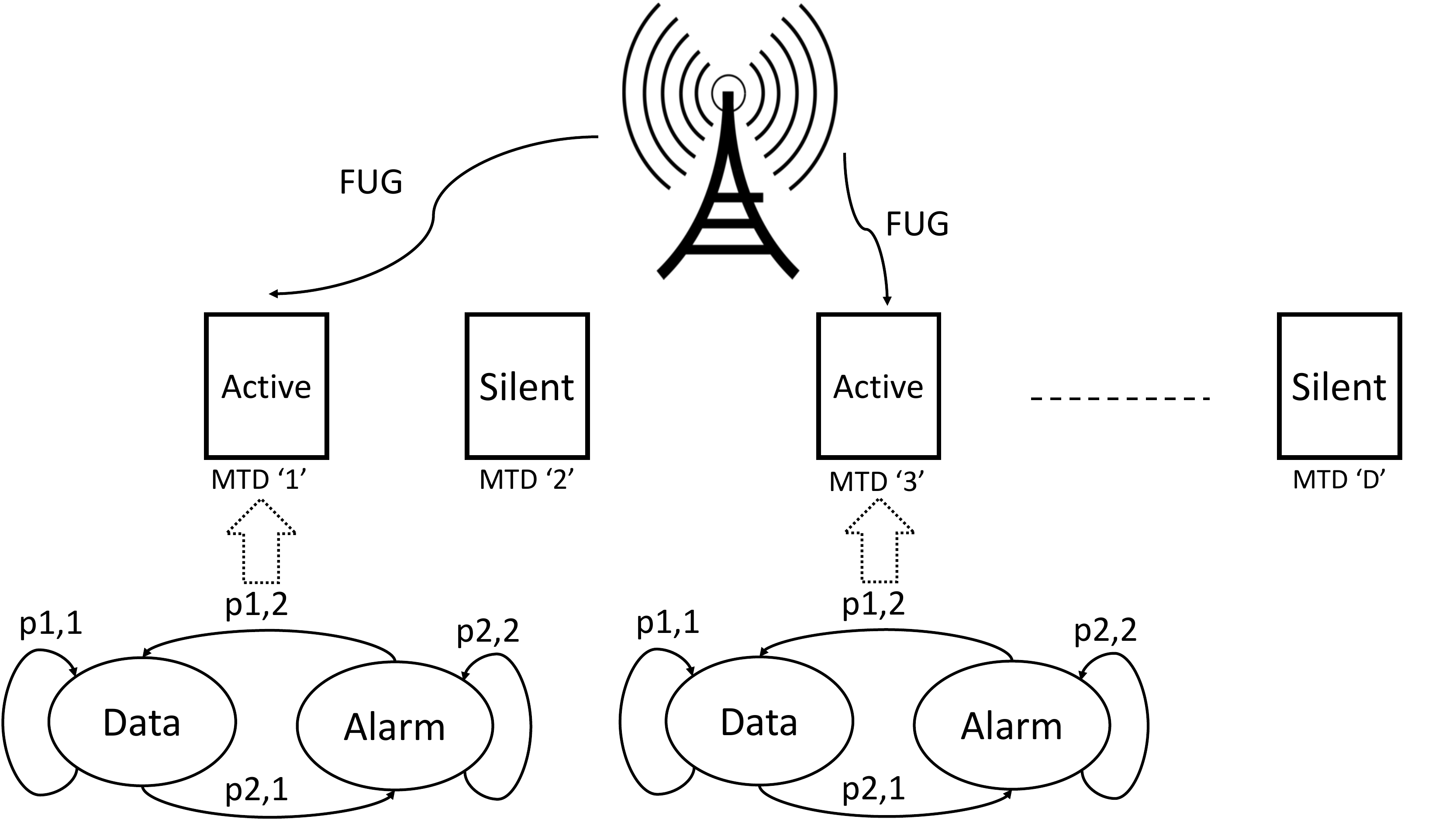}
\centering
\vspace{2mm}
\caption{The considered system model. A set of D devices, which can be active or silent. Active devices can be in either data or alarm state.}
\vspace{-0mm}
\label{System_Model}
\end{figure}

Consider a cellular network composed of a set $\mathcal{D}$ of static MTDs served by one BS with a limited number of frequency resources. The total number of devices is $D=|\mathcal{D}|$, where $|.|$ returns the length of a vector, such that each device $d_i \in \mathcal{D}$ can be either \textit{a) Active} or \textit{b) Silent}. When Active, devices can transmit either data packets with lower priority or alarm packets with higher priority. Each device $d_i \in D$ has fixed coordinate locations $x_i$ and $y_i$, which are known by the BS. In this work, for each one of the devices, we aim to first predict its state, namely silent or active, as well as the corresponding traffic priority. Thereafter, the serving BS schedules the set of active devices according to their priorities. We formulate the system model using an efficient traffic model called CMMPP \cite{6629817}. Then, we extend the baseline model to account for a more dense scenario using the M-CMMPP model, which consists of several background CMMPP processes. Fig.~\ref{System_Model} presents the system model.

%
\vspace{-0.0mm}
\subsection{CMMPP Traffic Model}
The Markov processes and Poisson processes are very popular traffic and queuing models \cite{1146393}, \cite{MILLER2012383}. Recently, the Markov Modulated Poisson Processes (MMPP) have been developed for traffic modeling scenarios, where a Poisson process with a rate $\lambda_i[t]$ changes according to the state of Markov chains $s_n[t]$, where $n$ is the number of MTDs. Consider a two-state MMPP, where the data state describes regular packet transmission, whereas the alarm state describes longer and higher priority packet transmission. Each device is transiting between these two states according to the respective probabilities based on the MMPP baseline model. Coupling Markov chains means that multiple chains mutually influence their transition probability matrices \cite{6629817}. Given the state transition probabilities, we build a state transition matrix $P_s$, while the state probability vector $\pi$ is defined using the respective state probabilities as follows: \vspace{-0mm}
\begin{align}\label{State_Matrix}
\begin{split}
P_s = 
\begin{pmatrix}
p_{1,1} & p_{1,2} & \cdots & p_{1,j}\\
p_{2,1} & p_{2,2} & \cdots & p_{2,j} \\
\vdots & \vdots & \ddots & \vdots \\
p_{i,1} & p_{i,2} & \cdots& p_{i,j}
\end{pmatrix},
\quad
\pi = 
\begin{pmatrix}
\pi_1 \\
\pi_2 \\
\vdots
\end{pmatrix}.
\end{split}
\end{align}

In addition, the MMPP-based source traffic can be simplified by considering only one background process to modulate all MTDs in case of a sudden event such as fire or high temperature. Thus, the background process influences all the MTDs in both space and time, but with different strengths according to their distances from the epicenter (position of the background process) and time of that background process. Moreover, it causes some devices to transition from state 1 (data) to state 2 (alarm), while others remain in their current state. The state probability matrix is composed of two sub-matrices, namely the coordinated matrix $P_C$ and the uncoordinated matrix $P_U$. The former describes the behaviour of the devices near to the event and its main characteristic is to issue an alarm at a time (alarm state), then go back to the data state again. On the other hand, the latter describes the behaviour of the devices away from the background process and its main characteristic is to remain at the data state and never switch to the alarm state. The background process generates samples $\theta_n[t]$, which are a function of time and space for all devices. The  state probability matrix is calculated as follows:
\begin{align}
\label{Transition_eq}
P_n[t] &= \theta_n[t] . P_C + (1 - \theta_n[t]) . P_U, \\
\label{theta_samples}
\theta_n[t] &= \delta_n . \theta[t],
\end{align}
where $\theta[t] \in [0,1]$ consists of samples, uniformly distributed over time, that describe how the devices are affected by the background process in time. $\delta_n$ follows standard normal distribution whose samples describe how the devices are affected by the background process in space. In addition, the mean and variance of $\delta_n$ describe the epicenter of the background process and how strong is the background process (high variance, means strong event and further devices are affected), respectively. Devices near the epicenter of the background process are highly affected by the process and transit to state 2 (alarm). Modeling $\delta_n$ as normal distribution describes the behaviour of sudden events (for example fire or high temperature) in real-time applications, where devices near to the center of the event have higher probabilities to be affected by that event than far devices. Multiplying $\theta[t]$ and $\delta_n$ results in $\theta_n[t]$, which describes how the devices are affected by the background process jointly in space and time. The same idea is repeated with the state probability vector $\pi_n[t]$, which will be composed of $\pi_C$ and $\pi_U$. Therefore, $\theta$ and $\delta_n$ are distributed as $\theta \sim \mathcal{U}(T, T+\tau)$ and $\delta_n \sim \mathrm{N}(\mathbf{\mu}, \mathbf{\Sigma})$, where $\mathcal{U}$ is the uniform distribution, $\mathrm{N}$ is the normal distribution, $T$ is the start instant of the background process, $\tau$ is the duration of the background process, while $\mathbf{\mu}$ and $\mathbf{\Sigma}$ represent the mean and the covariance matrix of the background process in x and y coordinates, respectively. Space and time distributions are independent. Therefore, the probability density function (PDF) of $\theta_n[t]$ is defined in \eqref{theta_n_t}, where $x,y \in \mathbb{R}_0^+$.
\vspace{-0.0mm}
\begin{align}\label{theta_n_t}
\begin{split}
f_{\theta_n}\! (x\!,\!y,t)\! =\! 
\begin{cases}
\!\frac{\frac{1}{\tau}}{2 \pi \sigma_x \!\sigma_y} \!\exp\!\left(\!-\frac{1}{2}\!\left[\!\left(\frac{x\!-\mu_x}{\sigma_x}\!\right)\!^2 \!+\! \left(\frac{y-\!\mu_y}{\sigma_y}\right)\!^2\right] 
\!\right)\\ 
\: \: \: \: \: \: \: \: \: \: \: \: \: \: \: \: \: \: \: \: \: \: \: \: \: \: \: \: \: \: \: \: \: \: \: \: \: \: \: \:  ,T\!<t<T+\tau; \\
0, \: \text{otherwise.}
\end{cases}  
\end{split}
\end{align}
\vspace{-0.0mm}

Simulating the CMMPP traffic model using the derived equations results in 4 plots describing the behaviour of MTDs in space and time as shown in Fig.~\ref{CMMPP_MCMMPP}. Fig.~\ref{CMMPP_MCMMPP} shows (a) the Startup state where all the devices are likely to transmit data; (b) the Data state in which some devices transmit data at different instants and using different packet lengths; (c) the Alarm state where devices near to the epicenter transmit alarm signal in a correlated behaviour with large packet lengths due to the activation of the background process; and finally (d) the Silent state where the devices that transmitted alarm tend to be silent \cite{6629817}.
\vspace{-0mm}
\subsection{M-CMMPP Traffic Model} \vspace{-0.0mm}
We introduced the CMMPP model based on one background process to represent a regular MTC use case. Some applications can have more than one background process at a time causing a bursty scenario and network congestion \cite{7996498}. Assume $M \geq 1$ background processes affect the transition of devices from the data state to the alarm state as follows:
\begin{align}\label{Theta_m}
\begin{split}
\theta_{n_m} [t] = \delta_{n_m}  \theta_m [t], \quad m = 1, 2, 3, ..., M.
\end{split}
\end{align}

As the background processes are independent, We start by calculating the probability of no alarms ($P_{NA}$), which is a function of the probability of alarm at a given process ($P_{m_{A}}$):
\begin{align}\label{P_NoAlarm}
\begin{split}
P_{NA} &= (1 - P_{1_{A}})  (1 - P_{2_{A}})  \cdots
 (1 - P_{M_{A}}) \\
& = \prod_{m=1}^{M} (1 - P_{m_{A}}), \quad m = 1, 2, 3, ..., M,
\end{split}
\end{align}
\begin{align}\label{P_Alarm_int}
P_{A}\! &= \! 1 \!-\! P_{NA} 
\!=\! 1 \!-\! \prod_{m=1}^{M} (1 \!-\! P_{m_{A}}), \quad m = 1, 2, 3, ..., M.
\end{align}
%
The overall $\theta_n [t]$, which is the probability of having alarms due to $M$ background processes, can be derived as follows:
\begin{align}\label{Theta_m_final}
\begin{split}
\theta_n [t] = 1 - \prod_{m=1}^{M} (1 - (\delta_{n_{m}} . \theta_m [t])).
\end{split}
\end{align}

Then, the overall $\theta_n [t]$ from \eqref{Theta_m_final} is used in \eqref{Transition_eq} to form the transition probability matrix. Therefore, $\theta_m$ and $\delta_{n_m}$ are distributed respectively as $\theta_m \sim \mathcal{U}(T_m, T_m+\tau_m)$ and $\delta_{n_m} \sim \mathrm{N}(\mathbf{\mu_m}, \mathbf{\Sigma_m})$, where $\mathcal{U}$ is the uniform distribution, $\mathrm{N}$ is the normal distribution, $\mathbf{\mu_m}$, $T_m$ is the start instant of the background process $m$, $\tau_m$ is the duration of the background process $m$, and $\mathbf{\mu_m}$ and $\mathbf{\Sigma_m}$ represent the mean and the covariance matrix of the background process $m$ in x and y coordinates, respectively. Space and time distributions are independent. Therefore, the overall $\theta_n[t]$ is defined as in \eqref{joint_n_t_theta_m} (on the top of the next page), where $x,y \in \mathbb{R}_0^+$, and $m = 1, 2, ..., M$.
%
\begin{figure*}
\begin{align}\label{joint_n_t_theta_m}
f_{\theta_n} (x,y,t) = 
& \begin{cases}
\frac{1}{\tau_m}\left[1-\left(\Pi_{n=1}^{M} \left(1 -\frac{\exp\left(-\frac{1}{2}\left[\left(\frac{x-\mu_{n_x}}{\sigma_{n_x}}\right)^2 +  \left(\frac{y-\mu_{n_y}}{\sigma_{n_y}}\right)^2\right] 
\right) }{2 \pi \sigma_{n_x} \sigma_{n_y}} 
\right) \right)\right], T_m<t<T_m+\tau_m;
\\
0, \text{otherwise.}
\end{cases} 
\end{align}
\hrule
\end{figure*}


\begin{figure*}[t!]
    \centering
    \subfloat[Startup]{\includegraphics[width=0.3\textwidth]{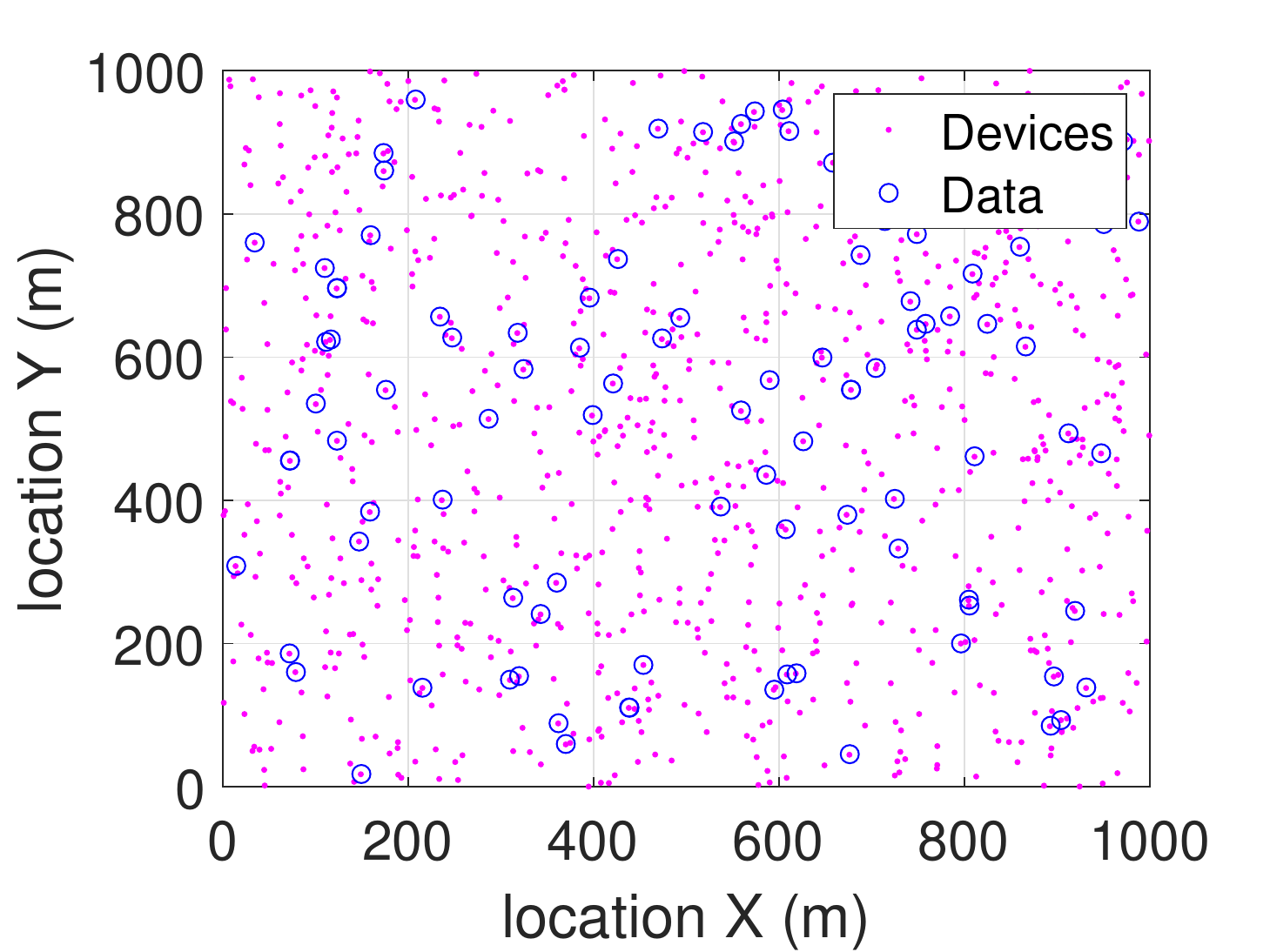}}
    \hskip -2.28ex
    \subfloat[Data]{\includegraphics[width=0.25\textwidth,trim={2.5cm 0 0 0},clip]{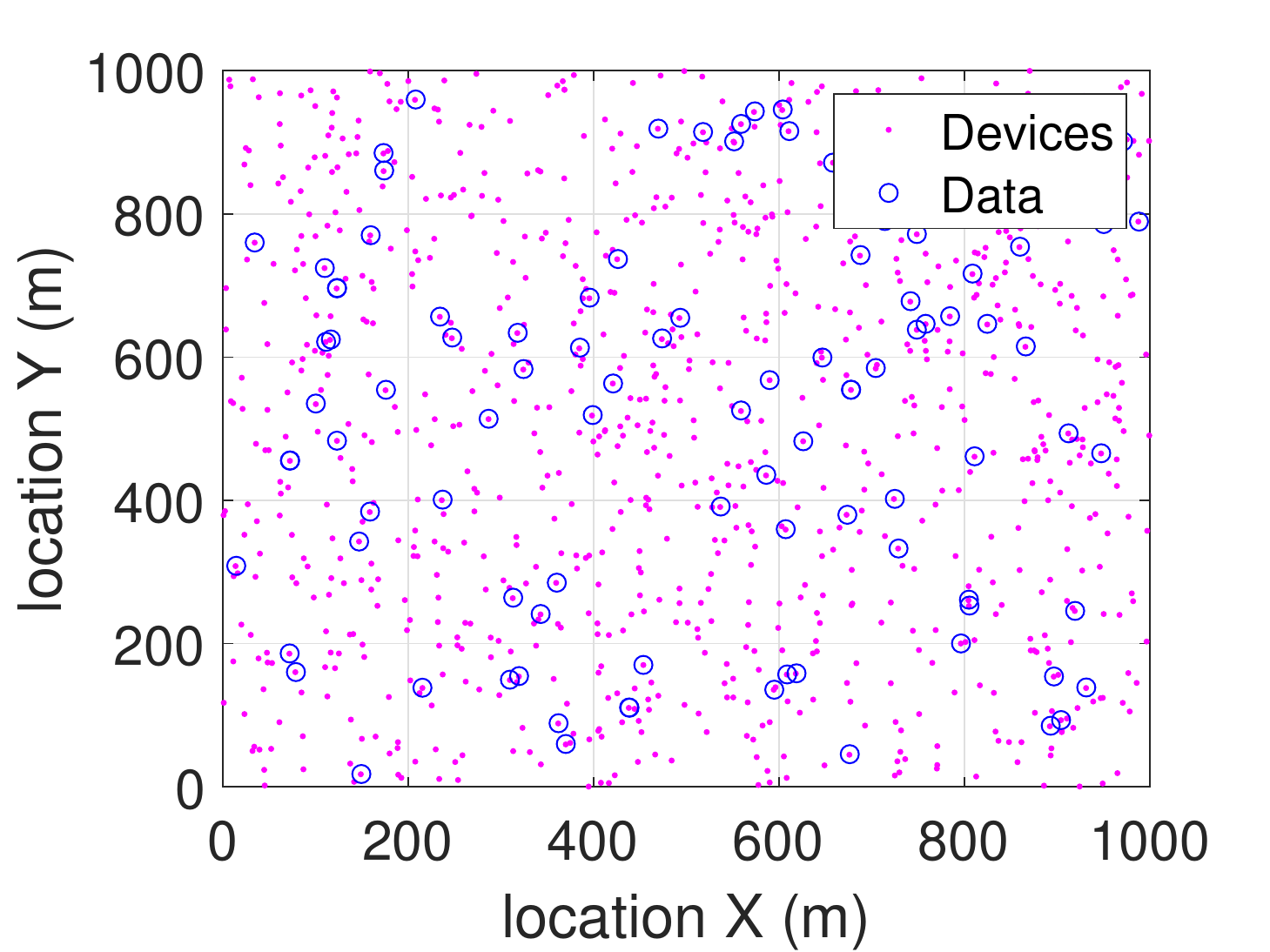}}
    \hskip -2.28ex
    \subfloat[Alarm]{\includegraphics[width=0.25\textwidth,trim={2.5cm 0 0 0},clip]{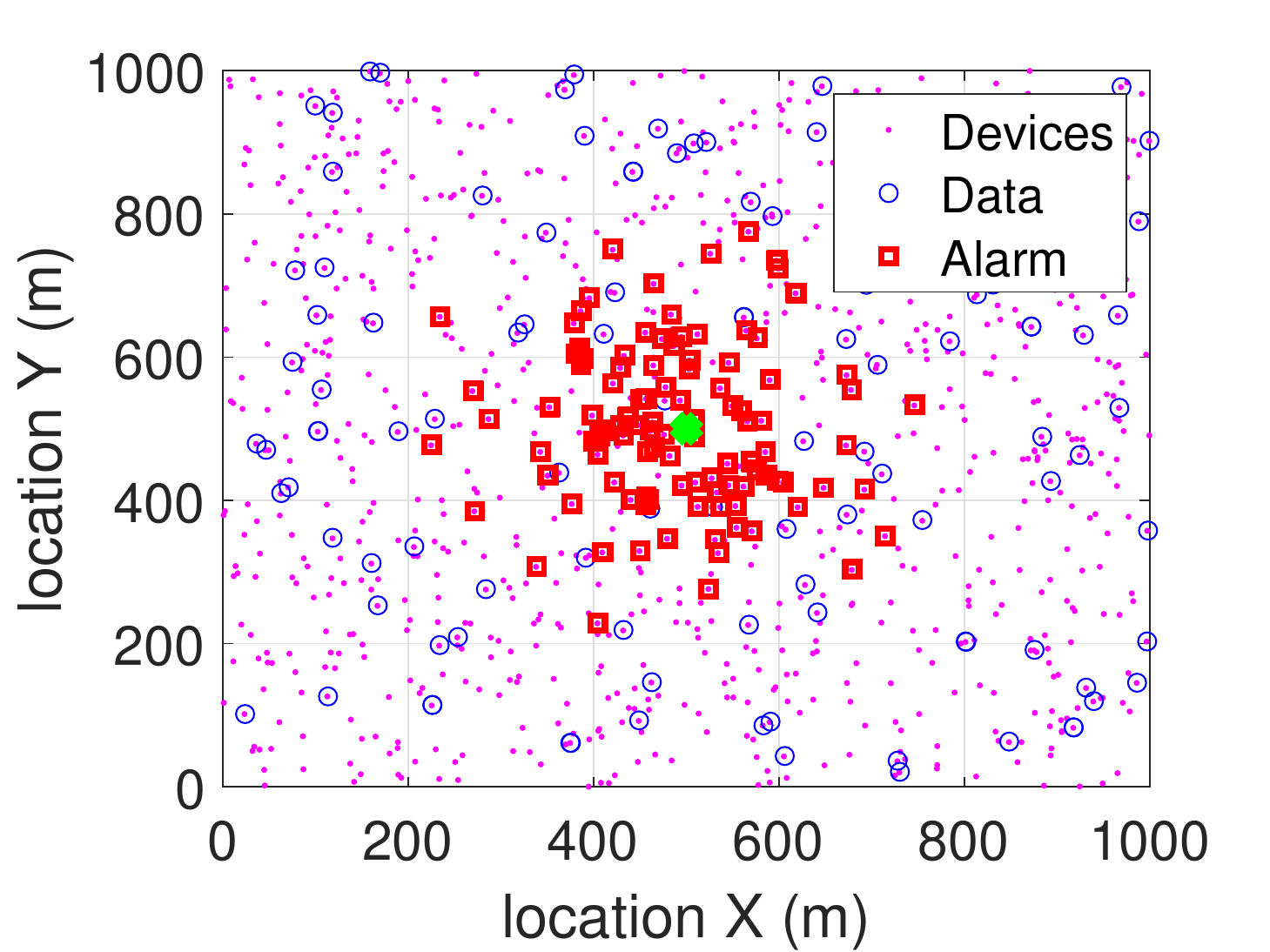}}
    \hskip -2.28ex
    \subfloat[Silent]{\includegraphics[width=0.25\textwidth,trim={2.5cm 0 0 0},clip]{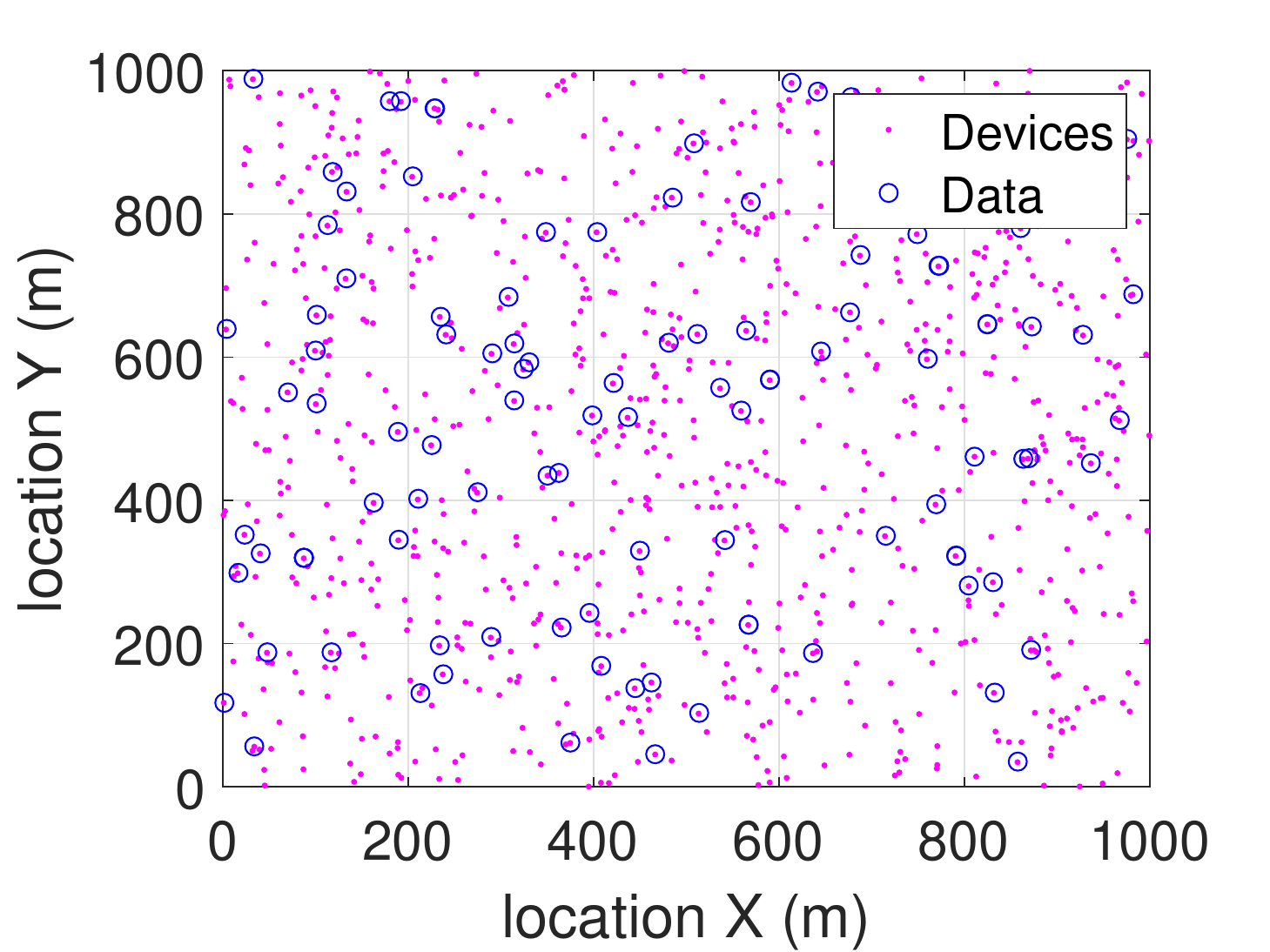}
    }
    \vspace{0.05ex}
    \subfloat[Startup]{\includegraphics[width=0.3\textwidth]{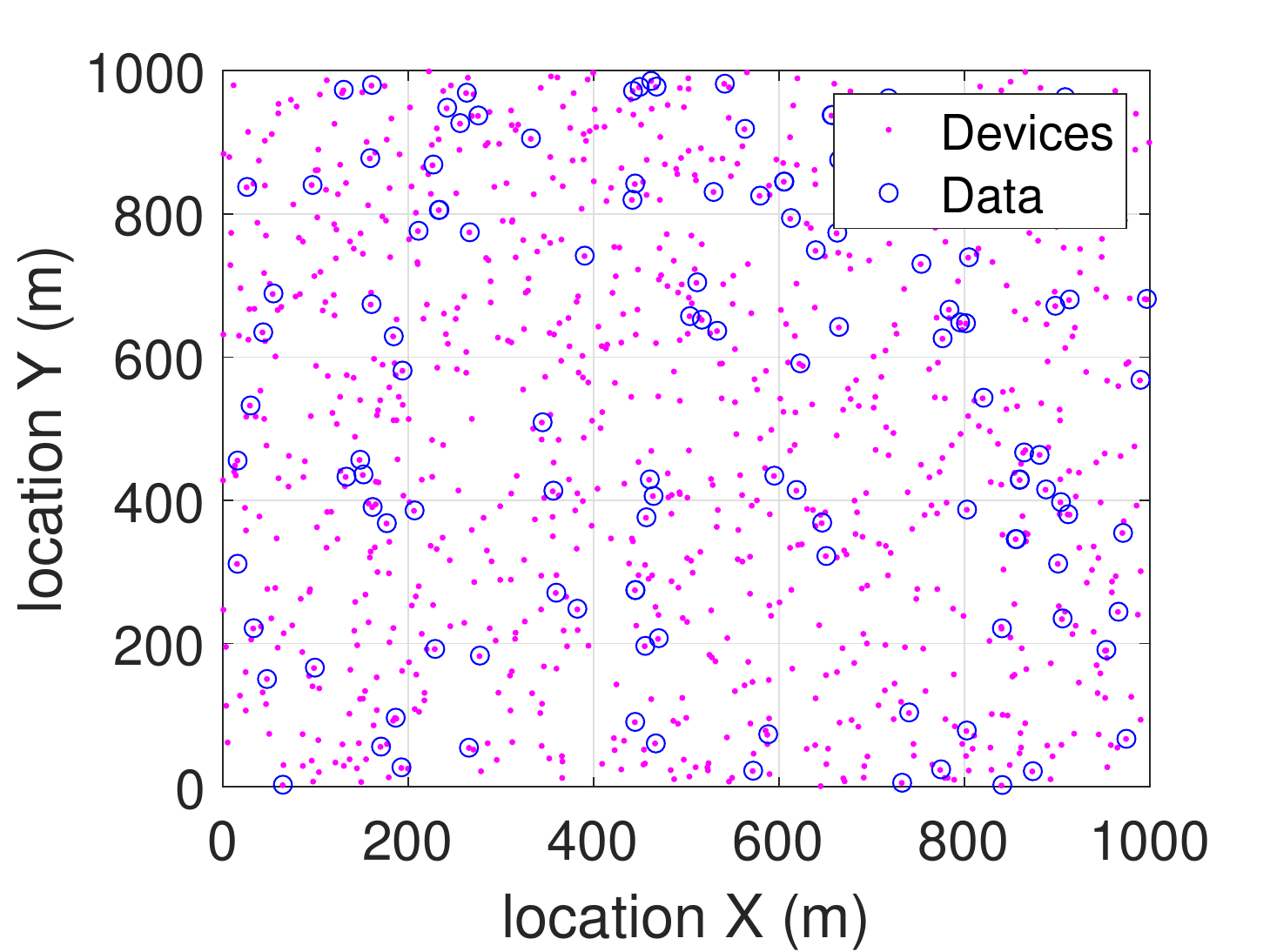}}
    \hskip -2.28ex
    \subfloat[Data]{\includegraphics[width=0.25\textwidth,trim={2.5cm 0 0 0},clip]{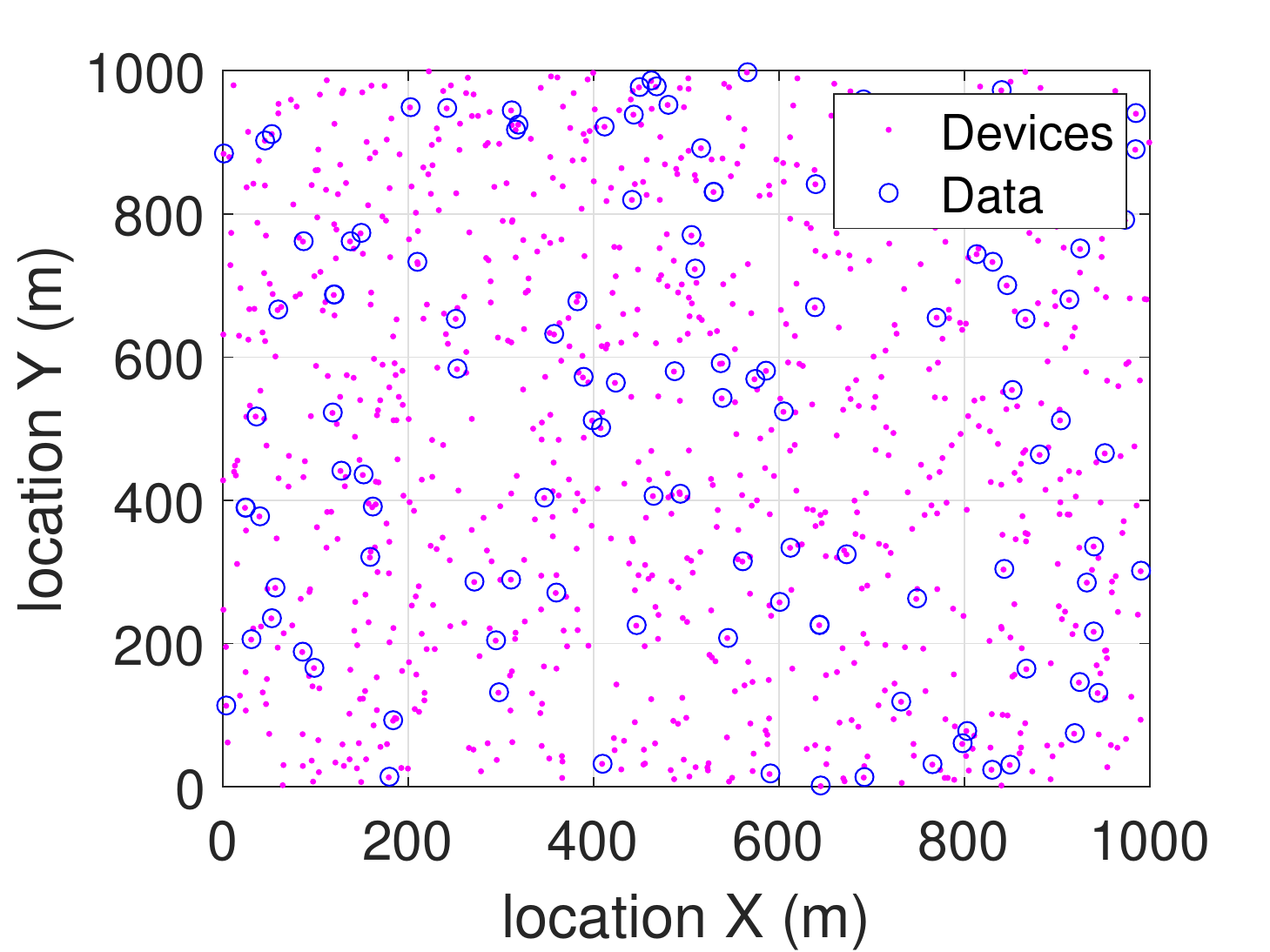}}
    \hskip -2.28ex
    \subfloat[Alarm]{\includegraphics[width=0.25\textwidth,trim={2.5cm 0 0 0},clip]{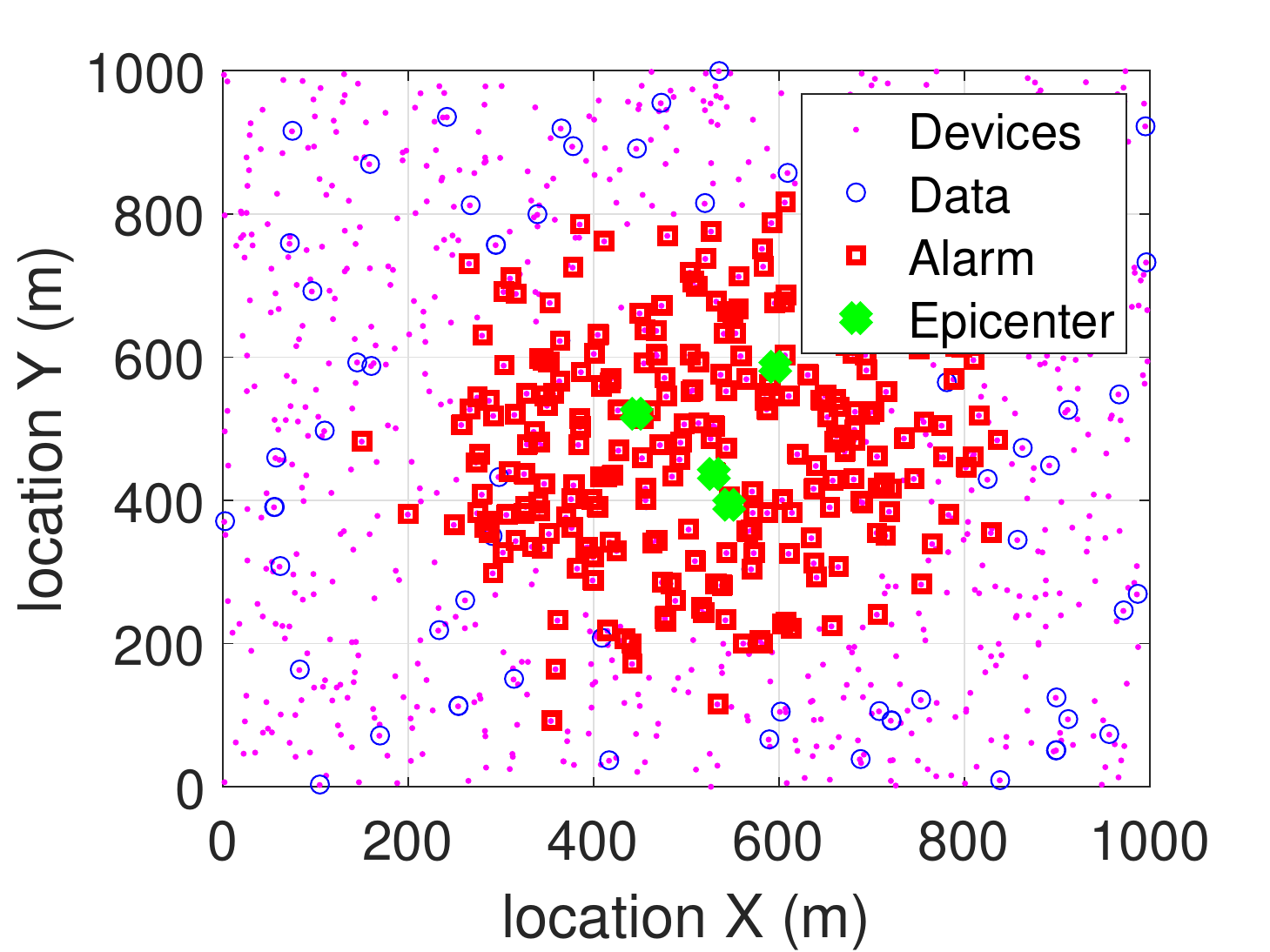}}
    \hskip -2.28ex
    \subfloat[Silent]{\includegraphics[width=0.25\textwidth,trim={2.5cm 0 0 0},clip]{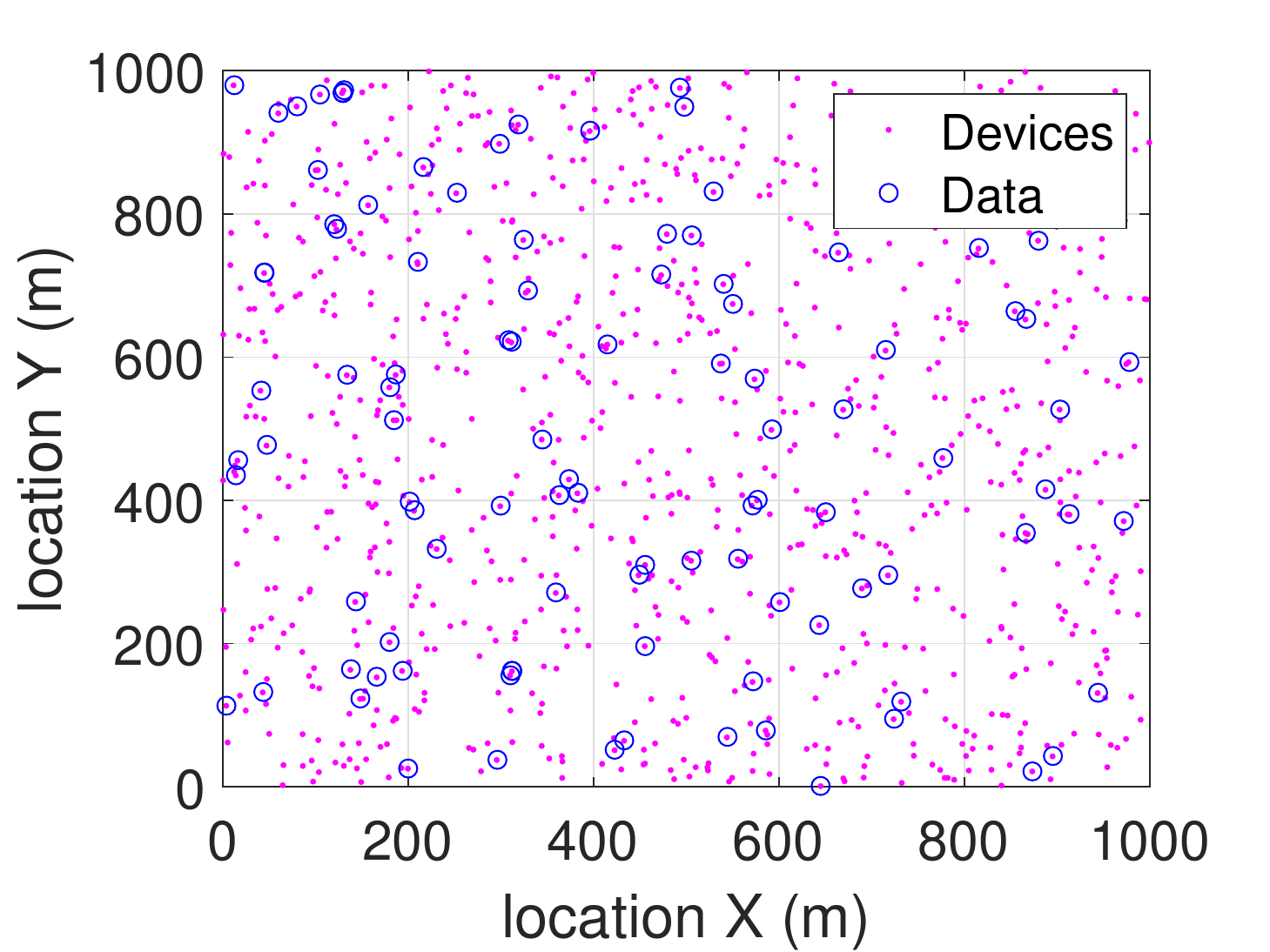}
    }
    \caption{Traffic modeling results: 1000 MTD, 60 s runtime, 1000 m $\times$ 1000 m area, $\lambda_{Data}$ = $\frac{1}{3600}$, $\lambda_{Alarm}$ = 1, and uniformly distributed processes epicenters, intervals, and variances. CMMPP results: (a) to (d). M-background processes CMMPP results (M = 4): (e) to (h).}
    \label{CMMPP_MCMMPP} \vspace{-0mm}
\end{figure*}

Simulating the proposed M-CMMPP traffic model using 4-background processes shows similar results to the original CMMPP model, but with an extremely larger number of data and alarm packets as shown in Fig.~\ref{CMMPP_MCMMPP} (e) to (h). This can cause network congestion, which is one of the most challenging problems in MTC applications while using current cellular network technologies. \vspace{-0.0mm}


\vspace{-0.0mm}
\section{The Proposed Fast Uplink Grant Algorithm}\label{FUG_alg_section}
\vspace{-0.0mm}
The proposed FUG model is divided into 3 main parts: \textit{i) device classification}, \textit{ii) traffic prediction}, and \textit{iii) resource allocation}.
Firstly, the BS predicts which devices, within a certain application under its coverage area, have a higher priority compared to other devices and need to be served first. The second step is estimating the time, which each device needs to access the network. Finally, the BS uses the results from the device classification and traffic prediction to allocate resources to the selected devices at specific time instants. The data collection, pre-processing, and computations done on it are performed on a server on the BS end, where it has better computational and energy resources that can limit the latency of the computations. \vspace{-0mm}




\subsection{Device Classification Using SVM}
According to the CMMPP model, the serving BS classifies devices into two groups. While devices in the first group transmit only data all the time, the second group transmit data and alarm. Devices that transmit alarms have higher priority than devices that transmit data. Specifically, our first step is a typical binary classification problem. Binary classification is a type of supervised learning, where a model is developed to classify some candidates among two classes. The first step in device classification is collecting an appropriate dataset with desired features and labels, that reflects different scenarios of the application for a period of time. 
Features are the input data that the classifier should take into consideration and learn their pattern and how they are related to their corresponding labels. In MTDs classification, features are position coordinates of the devices, while labels are the data or alarm classes. Usually, binary classification problems are considered as finding the best decision boundary that separates the classes correctly. 
\vspace{-0.0mm}
The optimum decision boundary can be found using SVM algorithm. Define training points $T = {(\vec{x}_i,z_i)}$, where $\vec{x}_i$ are the features, and $z_i$ are the labels $(-1,1)$. Then, define the decision hyperplane as $b$ and a normal vector \textbf{$\vec{w}$} perpendicular to the hyperplane. Finally, define a point $\vec{x}$ to be on the hyperplane, so that: \vspace{-0mm}
\begin{align}\label{Hyperplane}
\begin{split}
\vec{w}^T \vec{x} = -b. \vspace{-0mm}
\end{split}
\end{align}
The SVM aims to maximize the width between the nearest features from one class to the other (support vectors of each class) \cite{10.5555/1394399}. Fortunately, this optimization problem is convex, which can be solved using the Lagrange multiplier theorem with any quadratic programming: \vspace{-0mm}
\begin{align}\label{Lagrangian}
    L = \frac{1}{2} \| w \|^2 - \sum \alpha_i \left[z_i (\vec{w}. \vec{x}_i + b) - 1\right],
\end{align}
where $\alpha_i$ is the Lagrange multiplier. Classify class 1 if:
\begin{align}\label{Classifier}
\begin{split}
\sum \alpha_i . y_i . \vec{x_i} . \vec{x} + b \leq 0.
\end{split}
\end{align}
By inserting features (device coordinates) and labels (class 1 or class 2) into the classifier, we find the optimal boundary between the two classes. However, this classifier is only applicable to linearly separable feature points. The CMMPP classes are non-linearly separable. Thus, to avoid this problem we use transformation kernels ($\phi$) to map the feature points into higher dimensions, where they can be linearly separable: \vspace{-0.0mm}
\begin{align}\label{Transf.}
\begin{split}
\phi: \vec{x} \xrightarrow{} \phi(\vec{x}).
\end{split}
\end{align}
The radial basis function (RBF) maps the data into an infinite-dimensional Hilbert space \cite{10.5555/1394399}. It is very simple and fast to remap feature points using such kernel transformations; here, we employ the Gaussian RBF kernel given as:
\begin{align}\label{eq83}
\begin{split}
K(\vec{x},\vec{x'}) = e^{-\frac{(\vec{x} - \vec{x'})^2}{2 \sigma^2}}, 
\end{split}
\end{align}
where $\vec{x}$ and $\vec{x'}$ are two features. The RBF SVM can not only extract the pattern of devices efficiently but also classify them according to their priorities. \vspace{-0mm}


\subsection{Traffic Prediction Using LSTM}
After predicting the priorities of the devices, the serving BS needs to predict the activation time and silent time of these devices to implement an efficient resource scheduler. 
In a simple RNN architecture, input features, which are the past status of MTDs collected by the BS, are updated each instant $t$ and fed into an ordinary ANN, where hidden layers are connected to form a feedback path. 
In such RNNs, there are three types of weights, which should describe the dependency between instants: \textit{i) $w_{xh}$} represents weights from input features to hidden layers; \textit{ii) $w_{hh}$} corresponds to weights from hidden layers at time instant $t-1$ to hidden layers at time instant $t$; and \textit{iii) $w_{hy}$} yields the weights from hidden layers to output. Those weights are used in the underlying prediction as follows:
\begin{align}\label{RNN_tanh}
h_t &= \tanh(w_{hh} h_{t-1} + w_{xh} x_t),\\
\label{Output_RNN}
y_t &= w_{hy} h_t,
\end{align}
where $\tanh(\cdot)$ is the hyperbolic tangent activation function, $h_{t-1}$ is the previous hidden layer at $t-1$ result from the same recurrent equation, and $x_t$ is the input features vector at $t$. Afterwards, a loss function and an optimizer are applied to find the correct weights that describe the relation between inputs and outputs \cite{article_AdGrad}.

\begin{figure}
\centering
 \includegraphics[width=0.96\columnwidth]{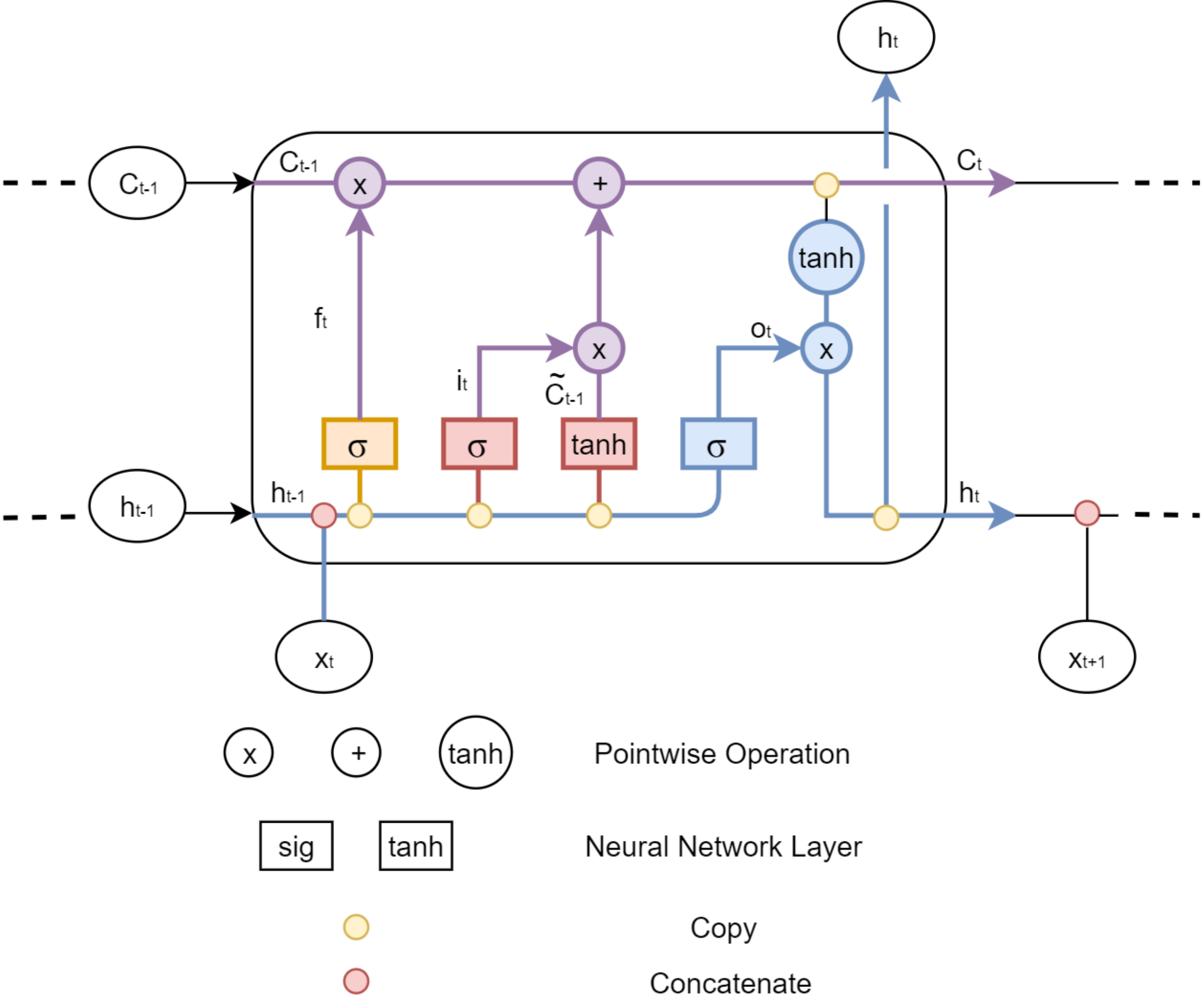} \vspace{1.5mm}
\caption{Long Short-Term Memory architecture.}
 \label{LSTM_Arch}
\end{figure}

Despite their astonishing ability to forecast the future, the RNNs still have a major weakness, namely, they have difficulty extracting relevant information located in the far past. Long sequences cause a major problem known as vanish gradient, where the relevant information is located far away in the past experiences almost zero gradient \cite{279181}. To solve the problem of long-term dependencies, Hochreiter and Schmidhuber introduced their LSTM architecture in 1997 \cite{article_LSTM}. It uses the same concept of basic RNN, but with complex four-gate functions connecting past and current instants to extract relevant information from long and short memories. As shown in Fig.~\ref{LSTM_Arch}, the LSTM has two inputs at each instant the short term and the long term memories given by $h_{t-1}$ and $C_{t-1}$, respectively. While the former yields the previous hidden layer just as in RNN, the latter allows the LSTM to learn what to add and what to ignore from the very long past to keep only relevant information for prediction. The LSTM can be classified into 4 main gates; \textit{1) Forget gate}, which describes accurately what information to forget from both input features and previous hidden state, \textit{2) Learn gate}, which is responsible for learning new features related to the model, \textit{3) Remember gate}, where the long term memory is updated, $C_t$, using the results of the forget gate and the learn gate, and \textit{4) Use gate}, where the current short term memory is updated. The LSTM key equations are presented as follows: \vspace{-0.0mm}
\begin{align}\label{LSTM_Vec1}
\begin{pmatrix} i_t \\ f_t \\ o_t \\ \tilde{C}_t^l \end{pmatrix} &= \begin{pmatrix} \sigma \\ \sigma \\ \sigma \\ tanh \end{pmatrix} \: W^l \: \begin{pmatrix} h_t^{l-1} \\ h_{t-1}^{l} \end{pmatrix}, \\
C_t^l &= f_t \: \odot \: C_{t-1}^l + i_t \: \odot \: \tilde{C}_t^l, \\
h_t^l &= o_t \: \odot \: tanh(C_t^l),
\end{align} \vspace{-0mm}
where $i_t$ is the output of the learn gate, $f_t$ is the output of the forget gate, $o_t$ is the output of the use gate, $\tilde{C}_t^l$ is the initial long term memory vector, $\sigma$ is the sigmoid function, $W^l$ is the weighs vector, $h_t^{l-1}$ is the current initial hidden layer vector, $h_{t-1}^{l}$ is the previous hidden layer vector, $C_t^l$ is the current updated long term memory, $\odot$ is a point-wise multiplication, and $h_t^l$ is the current updated hidden layer vector. To address the computation efficiency of the LSTM architecture, we formulate the spatial complexity of each LSTM layer in terms of the number of parameters ($K$) as follows \cite{karpathy2015visualizing}:
\vspace{-0mm}
\begin{align}\label{LSTM_Complex}
\begin{split}
K = 4 \, (I+1) \, O+O^2,
\vspace{-0mm}
\end{split}
\end{align}
where $I$ is the size of the input vector to an LSTM layer, and $O$ is the size of the output vector of an LSTM layer.

The serving BS needs to collect a relevant dataset, which contains the status of each MTD for a period of time. This dataset is then trained with an LSTM model to predict the status of the MTDs in the future for a certain period of time. The training and prediction phases should be continuously alternated. We divide the time axis into windows, each window consists of two phases: \textit{i)} training phase, and \textit{ii)} prediction and correction phase. In the training phase, the BS uses the collected dataset about each MTD activity for a certain period of time $T_{tr}$ to use it with the LSTM model to predict its activity in the following $T_{p}$. 
The prediction is subject to a degree of accuracy and therefore prone to errors. Particularly in this context, there are two types of errors: \textit{1)} the device is silent and it is predicted as active, and \textit{2)} the device is active and it is predicted as silent.

\begin{figure}[ht] \vspace{-0mm}
\centering
\includegraphics[width=0.97\columnwidth]{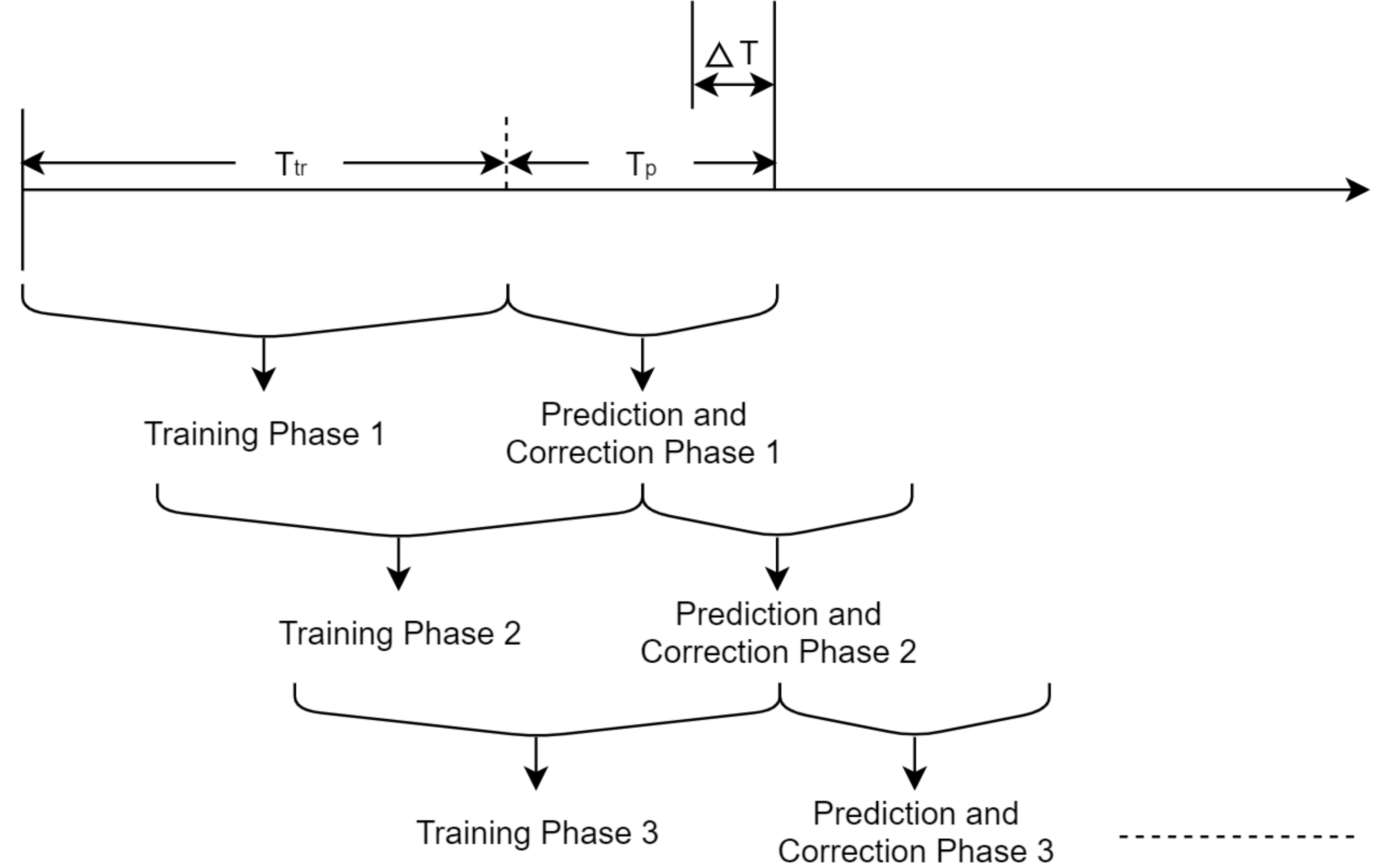}
\centering
\vspace{2mm}
\caption{Visualizing prediction of MTD status in real-time. Time axis is divided into windows, which are composed of training and prediction phases.}
\label{Sensor_Prediction} \vspace{-0mm}
\end{figure}

Assume the BS has feedback, which we will discuss later, that senses these errors and corrects them. Define $\Delta T$ as the correction time. The serving BS should correct errors, that exist in the first $T_{p} - \Delta T$ period of the predicted samples. Afterward, a new training phase starts by shifting the training window by $T_{p} - \Delta T$, where the BS uses the corrected prediction samples concatenated with the last $T_{tr} - (T_{p} - \Delta T)$ period of the previous training period to train new samples using the same model. Then, a prediction will be performed for the next $T_{p}$ with correction of errors for first the $T_{p} - \Delta T$ interval of predicted samples. The process repeats so that we have error-free data at every training phase. The proposed algorithm is illustrated in Fig.~\ref{Sensor_Prediction}. \vspace{0mm}


\subsection{Resource Allocation}

As the serving BS classifies the type of each MTD and predicts their traffic, the next step is to schedule the resources for these devices. The resource allocation algorithm is illustrated in Fig.~\ref{FUG_RES_ALL}. The BS predicts the active devices at each instant, then classifies their priorities and grants them the needed resources with given order according to their priorities. Devices from the same class are scheduled as first come first served (FCFS). In addition, the BS performs some error correction techniques to perform an accurate prediction without accumulation of errors at each phase. The serving BS implements different procedures to carry out feedback and error correction as follows:
\begin{enumerate}
    \item The serving BS avoids eventual prediction errors by adding safety margins. The length of $T_{m}$ is chosen according to an exploration rate\footnote{The exploration rate $\alpha$ controls the margin time $T_{m}$, available resources for RA, and available resources for random allocation. Adjusting different exploration rates is out of the scope of this work, therefore, we arbitrarily set these parameters based on experimentation.} $\alpha$. 
    \item When the MTD is predicted to be active, while it is silent, the BS senses that the allocated resource has not been used. Afterward, the status of this device is changed from active to silent in the dataset to avoid wrong future predictions. 
    \item When the MTD is predicted to be silent, while it is active, the device should wait for a period of $T_{m}$ to get a resource. If it does not get a resource, it should communicate with the BS using the RA procedure. Hence, the BS should dedicate some resources for the RA procedure to be used in case of prediction errors. The percentage $a$ of dedicated RA resources is also adjusted according to the exploration rate $\alpha$.
    \item The BS explores random devices, other than those that it has predicted, using available unused resources randomly according to $\alpha$. \vspace{-1mm}
\end{enumerate}
Algorithm \ref{alg1} summarizes the proposed fast uplink grant resource allocation learning algorithm including the discussed error correction techniques.

\tikzstyle{intt}=[draw,text centered,minimum size=6em,text width=5.25cm,text height=0.34cm]
\tikzstyle{intl}=[draw,text centered,minimum size=2em,text width=2.75cm,text height=0.34cm]
\tikzstyle{int}=[draw,minimum size=0.8em,text centered,text width=1.2cm]
\tikzstyle{inttt}=[draw,minimum size=0.3em,text centered,text width=1.8cm]
\tikzstyle{intg}=[draw,minimum size=3em,text centered,text width=6.cm]
\tikzstyle{dec}=[draw,shape=diamond, minimum size=0.2em,text centered,text width=0.8cm]
\tikzstyle{sum}=[draw,shape=circle,inner sep=2pt,text centered,node distance=3.5cm]
\tikzstyle{summ}=[draw, shape=trapezium,trapezium left angle=70,trapezium right angle=-70, inner sep=1.5pt,text centered,node distance=2.cm]
\tikzstyle{outt}=[coordinate,node distance=1.7cm]

\begin{figure}[t!]
\centering
     \begin{tikzpicture}[font=\footnotesize
       ,
       auto
     ]
       \node [summ] (st)  {Start, t=0};
       \node [int] (ti)  [node distance=0.25cm and -1cm,below =of st] {t=t+1};
       \node [dec] (kp)  [node distance=0.3cm and -1cm,below =of ti] {Still Error?};
       \node [inttt]  (ki1) [node distance=0.07cm and 1cm,below left=of kp] {Predict Traffic};
       \node [int]  (ki2) [node distance=0.07cm and 1cm,below right=of kp] {Use RA};
       \node [inttt]  (ki5) [node distance=0.22cm and -1cm,below =of ki2] {Predict Traffic};
       \node [int] (ki3) [node distance=2.15cm,below of=kp] {Prioritize};
       \node [int] (ki4) [node distance=0.68cm,below of=ki3] {Allocation};
       \node [dec] (ki6)  [node distance=0.3cm and -1cm,below =of ki4] {Error?};
       \node [inttt]  (ki7) [node distance=0.6cm and 0.7cm,right =of ki6] {Wait Margin};
       \node [outt]  (ki8) [node distance=1.7cm and 0.3cm, right=of ki7] {};
       \node [outt]  (ki9) [node distance=1.7cm and 3.06cm, right=of ti] {};
       \node [outt]  (ki10) [node distance=1.7cm and 3.18cm, left=of ki6] {};
       \node [outt]  (ki11) [node distance=1.7cm and 3.2cm, left=of ti] {};
       
       \draw[->] (st) -- (ti);
       \draw[->] (ti) -- (kp);
       \draw[->] (kp) -- ($(kp.west)+(0,0)$) -| (ki1) node[above,pos=0.15] {No} ;
       \draw[->] (kp) -- ($(kp.east)+(0,0)$) -| (ki2) node[above,pos=0.15] {Yes};
       \draw[->] (ki1) |- (ki3);
       \draw[->] (ki2) -- (ki5);
       \draw[->] (ki5) |- (ki3);
       \draw[->] (ki3) -- (ki4);
       \draw[->] (ki4) -- (ki6);
       \draw[->] (ki6) -- (ki7) node[above,pos=0.15] {Yes};
       \draw[-] (ki7) -- (ki8);
       \draw[-] (ki8) -- (ki9);
       \draw[->] (ki9) -- (ti)node[above,pos=0.25] {};
       \draw[-] (ki6) -- (ki10)node[above,pos=0.15] {No};
       \draw[-] (ki10) -- (ki11);
       \draw[->] (ki11) -- (ti)node[above,pos=0.25] {};
     \end{tikzpicture}
     \caption{Fast uplink grant resource allocation algorithm.}  \vspace{-0mm}
     \label{FUG_RES_ALL} 
\end{figure}
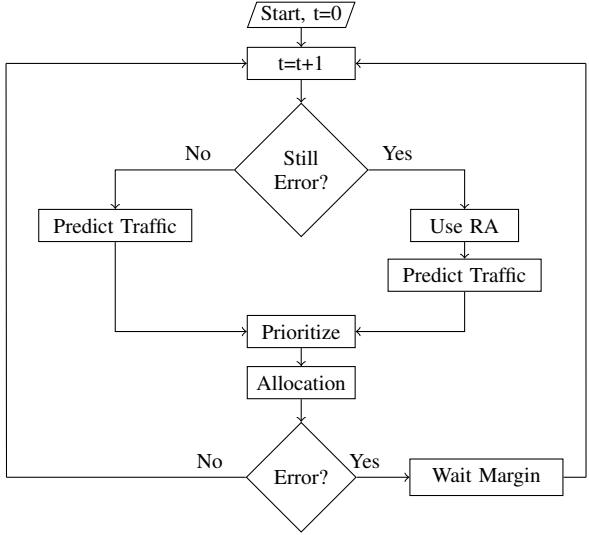

\begin{algorithm}[]
\SetAlgoLined
Define $T_{tr}$.

Define $T_{p}$.

Define $\Delta T$.

Define exploration rate $\alpha$.

\While{True}{
    Run device classification.
    
    Run traffic prediction.
    
    Prioritize active devices.
    
    Allocate resources.
    
    \If{error}{
        Wait $T_m$.
        
        \If{still error}{
             Use RA.
        }
        
    }
    
    Explore random devices.
    
    Correct errors in $T_p$ - $\Delta T$.
    
    Create new error-free training sequence.
    }
\caption{Fast uplink grant resource allocation algorithm.}
\label{alg1}
\end{algorithm}

\vspace{-0.0mm}

\section{Results and Discussion}\label{results_disc}
In this section, we present the simulation results of the proposed FUG. First, we introduce the performance evaluation metrics. Afterward, the SVM classification results are presented and compared to other classifiers. Then, we present the LSTM prediction of sensors activity. Finally, we apply those results along with the error correction techniques to schedule resources to MTDs. The proposed FUG model is compared with grant-based RA (GB-RA), random FUG, where the serving BS randomly allocates resources to devices, and genie-aided FUG, where BS knows perfectly the traffic of each device and its priority. This comparison is done while adjusting the number of available frequency resources dedicated for 1000 devices in a 1000 m $\times$ 1000 m deployment area within 60 seconds. Moreover, The exploration rate $\alpha$ is set to 0.1, a = $\alpha$, and $T_m = 20 \: ms$. The simulations are carried out on python using Keras library. In addition, we use a single NVIDIA Tesla V100 GPU and 10 GB of RAM on a Linux operating system server dedicated for the researchers in the Center of Wireless Communications (CWC) at the University of Oulu to train the model.
\vspace{-0.0mm}

\subsection{Performance Evaluation Metrics}

Both the device classification and traffic prediction algorithms are considered binary classification problems. Thus, there are many appropriate evaluation metrics, which are suitable for skewed data applications (rare alarms) such as
\cite{book_ML}, \cite{SOKOLOVA2009427}:
\begin{itemize}
    \item \textbf{The confusion matrix} is the most important evaluation method, which illustrates the number of correct and wrong classifications in each class.
    \item \textbf{The classification accuracy} describes the overall performance of the classifier.
    \item \textbf{The precision ($P$) and recall ($R$).} The former describes the ratio of true predicted samples for each class to the total predicted samples of that class, while the latter describes the ratio of true predicted samples for each class to the total actual samples of that class.
    \item \textbf{The f1-score ($f1s$)} combines the $P$ and $R$ measurements via harmonic mean, resulting in a percentage near to the minimum between them.
    \begin{align}\label{f1_Score}
        f1s = \frac{2 \, R \, P}{R + P}.
    \end{align}
\end{itemize}

We evaluate the performance of the network with respect to throughput. The transmission rate is calculated using 
$C = \log (1 + \mathsf{SNR} \: |h|^2 )$,
where $\mathsf{SNR}$ is the signal to noise ratio, and $h$ is the given channel coefficient between an MTD and the BS. Each MTD will have a certain rate depending on the SNR and the channel condition between that device and the serving BS. In addition, for each transmission, different rates exist depending on the transmission power of an MTD and channel condition at the time of transmission. As our main scope is to compare FUG to other allocation techniques, we assume the radio channel to be degraded by flat fading.

Communication systems have different sources of delay such as hardware delay $T_h$, queuing delay $T_q$, and transmission delay $T_t$. The access delay, $T_a$, is the time difference between the moment an MTD is ready for transmission and the moment it receives a resource. It is a function of hardware delay, signaling overhead delay, and queuing delay originated from the existence of a lower number of resources compared to the number of ready devices at a time. In our simulation, we neglect the hardware delay and focus only on the access delay as a function of queuing delay and message exchange between MTDs and the BS. The access delay can be calculated as follows:
\begin{align}\label{DELAY1}
T_{total} &= T_h + T_{overhead} + T_q + T_t,\\
\label{DELAY2}
T_{a} &= T_{h} + T_{overhead} + T_q = T_{total} - T_t.
\end{align}

\subsection{Simulation Results}
\subsubsection{Device Classification}
We initially collect the training set by running multiple CMMPP and M-CMMPP simulations with uniformly distributed epicenters, variances, and intervals of background processes. Then, we pre-process the data by balancing, normalizing and then removing redundancies to ease the training phase and extract the correct features. Afterward, we compare different classification algorithms to a new CMMPP/M-CMMPP model as shown in table \ref{Dev_Class}. We observe that a polynomial kernel (degree = 6) SVM provides good performance in the CMMPP case, but fails in the network congestion case. An ANN architecture with 4 hidden layers (16, 32, 8, and 4 neurons) works very well in both cases in terms of data and alarm classification, where it introduced the lowest errors (only 12 alarm errors in case of CMMPP and 2 alarm errors in case of M-CMMPP) in classifying alarms compared to all classifiers. However, it has a relatively large number of errors in classifying data devices.

The radial basis function SVM, decision trees (DT), and random forests (RF) produce better results than ANN in classifying data devices and almost similar results as ANN in classifying alarm devices. The radial basis function SVM is the simplest algorithm among them and works extremely fast. It produces a recall of 0.87 and 0.88 for data and alarm classification, respectively in the case of CMMPP traffic and a recall of 0.96 and 0.97 for data and alarm classification in the case of M-CMMPP traffic. These small number of errors reflect the strength of the RBF SVM classifier in prioritizing the devices. In addition, the RBF SVM needs less than 3 seconds of execution time for both training and prediction, which reflects its high efficiency. Hence, RBF SVM is the chosen algorithm to be used by the BS to classify devices.

\begin{table}[!t]
\centering
    \caption{Confusion matrix, accuracy, precision, recall and f1-score for CMMPP device classification (support: 544 data and 113 alarm) and M-CMMPP device classification (support: 539 data and 153 alarm).}
	\label{Dev_Class}
\begin{tabular}{|l|l|l|l|l|l|}
\hline
 & Algorithm & Conf. Mat. & Acc. & $P$\&$R$ & $f1-Sc.$\\ \hline
\multirow{10}{*}{\rotatebox[origin=c]{90}{CMMPP}} & Poly. SVM & \begin{tabular}{@{}ll@{}}
                            354 & 190\\
                            7 & 106\\
                            \end{tabular} & 0.90 & 
                           \begin{tabular}{@{}ll@{}}
                           0.98  & 0.65\\
                           0.36 & 0.94\\
                         \end{tabular}  & \begin{tabular}{@{}l@{}}
                                           0.78\\
                                           0.52\\
                                           \end{tabular}
\\ \cline{2-6}
 & RBF SVM & \begin{tabular}{@{}ll@{}}
                            474 & 70\\
                            14 & 99\\
                            \end{tabular} & 0.87 & 
                            \begin{tabular}{@{}ll@{}}
                           0.97  & \textbf{0.87}\\
                           0.59 & \textbf{0.88}\\
                         \end{tabular}  & \begin{tabular}{@{}l@{}}
                                           0.92\\
                                           0.70\\
                                          \end{tabular}
\\ \cline{2-6}
 & ANN & \begin{tabular}{@{}ll@{}}
                            462 & 82\\
                            12 & 101\\
                            \end{tabular} & 0.86 & \begin{tabular}{@{}ll@{}}
                           0.97  & 0.85\\
                           0.55 & 0.89\\
                         \end{tabular}  & \begin{tabular}{@{}l@{}}
                                           0.91\\
                                           0.68\\
                                          \end{tabular}
\\ \cline{2-6}
 & DT & \begin{tabular}{@{}ll@{}}
                            478 & 66\\
                            17 & 96\\
                            \end{tabular} & 0.87 & \begin{tabular}{@{}ll@{}}
                           0.97  & 0.88\\
                           0.59 & 0.85\\
                         \end{tabular}  & \begin{tabular}{@{}l@{}}
                                           0.92\\
                                           0.70\\
                                          \end{tabular}
\\ \cline{2-6}
 & RF & \begin{tabular}{@{}ll@{}}
                            473 & 71\\
                            15 & 98\\
                            \end{tabular} & 0.87 & \begin{tabular}{@{}ll@{}}
                           0.97  & 0.87\\
                           0.58 & 0.87\\
                         \end{tabular}  & \begin{tabular}{@{}l@{}}
                                           0.92\\
                                           0.70\\
                                          \end{tabular}
\\ \Xhline{2\arrayrulewidth}

\multirow{10}{*}{\rotatebox[origin=c]{90}{M-CMMPP}} & Poly. SVM & \begin{tabular}{@{}ll@{}}
                            539 & 0\\
                            153 & 0
                            \end{tabular} & 0.85 & 
                           \begin{tabular}{@{}ll@{}}
                           0.78  & 1.00\\
                           0.00 & 0.00\\
                         \end{tabular}  & \begin{tabular}{@{}l@{}}
                                           0.88\\
                                           0.00\\
                                           \end{tabular}
\\ \cline{2-6}                               
 & RBF SVM & \begin{tabular}{@{}ll@{}}
                            518 & 21\\
                            5 & 148
                           \end{tabular} & 0.97 &
                            \begin{tabular}{@{}ll@{}}
                           0.99  & \textbf{0.96}\\
                           0.88 & \textbf{0.97}\\
                         \end{tabular}  & \begin{tabular}{@{}l@{}}
                                          0.97\\
                                           0.92\\
                                          \end{tabular}
\\ \cline{2-6}  
 & ANN & \begin{tabular}{@{}ll@{}}
                            486 & 53\\
                            2 & 151
                            \end{tabular} & 0.92 & \begin{tabular}{@{}ll@{}}
                           1.00  & 0.90\\
                           0.74 & 1.00\\
                         \end{tabular}  & \begin{tabular}{@{}l@{}}
                                           0.95\\
                                           0.85\\
                                          \end{tabular}
\\ \cline{2-6}  
 & DT & \begin{tabular}{@{}ll@{}}
                            518 & 21\\
                            8 & 145
                            \end{tabular} & 0.96 & \begin{tabular}{@{}ll@{}}
                           0.98  & 0.96\\
                           0.87 & 0.95\\
                         \end{tabular}  & \begin{tabular}{@{}l@{}}
                                           0.97\\
                                           0.91\\
                                          \end{tabular}
\\ \cline{2-6}  
 & RF & \begin{tabular}{@{}ll@{}}
                            517 & 22\\
                            6 & 147
                            \end{tabular} & 0.96 & \begin{tabular}{@{}ll@{}}
                           0.99  & 0.96\\
                           0.87 & 0.96\\
                         \end{tabular}  & \begin{tabular}{@{}l@{}}
                                           0.97\\
                                           0.91\\
                                          \end{tabular}\\
\hline                              
\end{tabular} \vspace{-0mm}
\end{table}
\vspace{-0.0mm}
\subsubsection{Traffic Prediction} \vspace{-0.0mm}
The Numenta Anomaly Benchmark (NAB) is a time-series dataset, which contains 58 time-series data files designed to help researchers in time-series prediction and streaming anomaly detection applications \cite{7424283}. This dataset provides real-time data collected from sensors monitoring different physical quantities in industrial deployment scenarios. We do some pre-processing steps, where the sensors are active and need resources when their measurement exceeds a certain threshold. 
After pre-processing steps, 2 months of training data have been prepared to be used for the first training phase. The prediction and correction window is set as $T_p=10$ minutes and $\Delta T=5$ minutes for each iteration. To illustrate the performance of our scheme, we perform 4 prediction and correction iterations\footnote{For ease of implementation, we perform only 4 iterations to predict 40 minutes of activation. Note that, these prediction iterations could be extended deep into the future for hundreds or even thousands of iterations with almost the same performance thanks to the feedback and error correction procedures that are presented in Section \ref{FUG_alg_section}.C.}.

We build up an LSTM architecture with 2 hidden layers (150 and 100 Neurons), $20\%$ dropout, mean square error (MSE) loss function, 50 unrolling (create an array of 50 samples, then, for the next array, add one element and use the last 49 from the previous array), 50 epochs, and using the Adam optimizer \cite{kingma2017adam}. In our predictions, this architecture produces overall accuracy of $95\%$, the predictor failed 11 times to correctly infer the sensor activity of a total of 171 activation instants, and it wrongly predicts 41 times that the sensor is active, while it is silent. Furthermore, it achieves f1-Scores of 0.98 for silent prediction and around 0.90 for active predictions. According to \eqref{LSTM_Complex}, the number of parameters in our architecture is 2551100, which is computationally efficient in terms of spatial efficiency as the larger the number of parameters, the better the architecture can extract relevant information \cite{karpathy2015visualizing}. In addition, each training phase needs around 200 seconds of execution time.
We should point out that much deeper LSTM architecture would enhance the prediction accuracy as it may extract more relevant information, but it would increase complexity and time of training and prediction as well. In addition, the length of the training phase can be increased, at expense of a long time to extract the pattern. Selecting the appropriate neural network architecture is a very challenging research problem, where large architectures consume a long time, whereas short ones may introduce lower accuracy \cite{6152147}. The depth of the model and the length of the training data should be adjusted according to the available hardware at the BS, the criticality of the application, the number of devices, and the number of expected errors. The resulting errors are corrected, as mentioned, to have clean training data again to be used in the next phase. \vspace{-0mm} 




\subsubsection{Resource Allocation}

\begin{figure}[!t]
\centering
\includegraphics[trim=0 0 0 3mm, width=0.99\columnwidth]{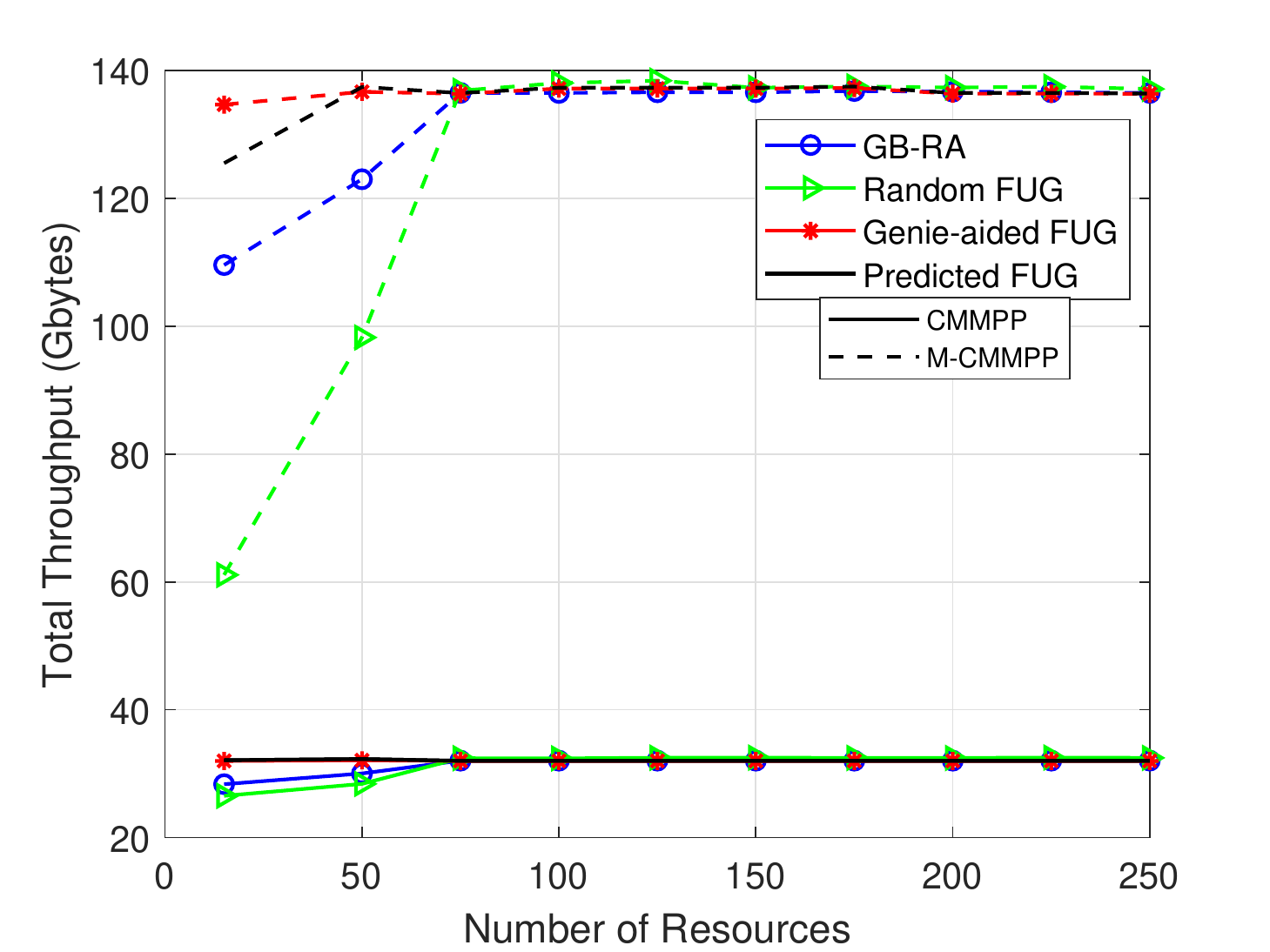}
\vspace{-0mm}
\centering
\caption{Total throughput while adjusting the number of available resources: 60 s runtime, 1000 MTDs, $\alpha$ = 0.1, $a$ = 0.1, and $T_m = 20 \: ms$. Beyond 75 frequency resources, all schemes converge.}
\label{Throughput}
\vspace{-1mm}
\end{figure}

Throughput and access delay are monitored for different allocation schemes while adjusting the number of available frequency resources at the BS. In Fig. \ref{Throughput}, the throughput, which is the total successfully received packets, is plotted while adjusting the number of available resources during 60 seconds. Random FUG resource allocation has the worst performance, whereas the predicted FUG outperforms the GB-RA and almost achieves the genie-aided FUG for both CMMPP and M-CMMPP. We notice that as the number of available resources increases beyond 75 frequency resources, all schemes start to converge and perform well.

\begin{figure}[!t]
\centering
\includegraphics[width=0.981 \columnwidth]{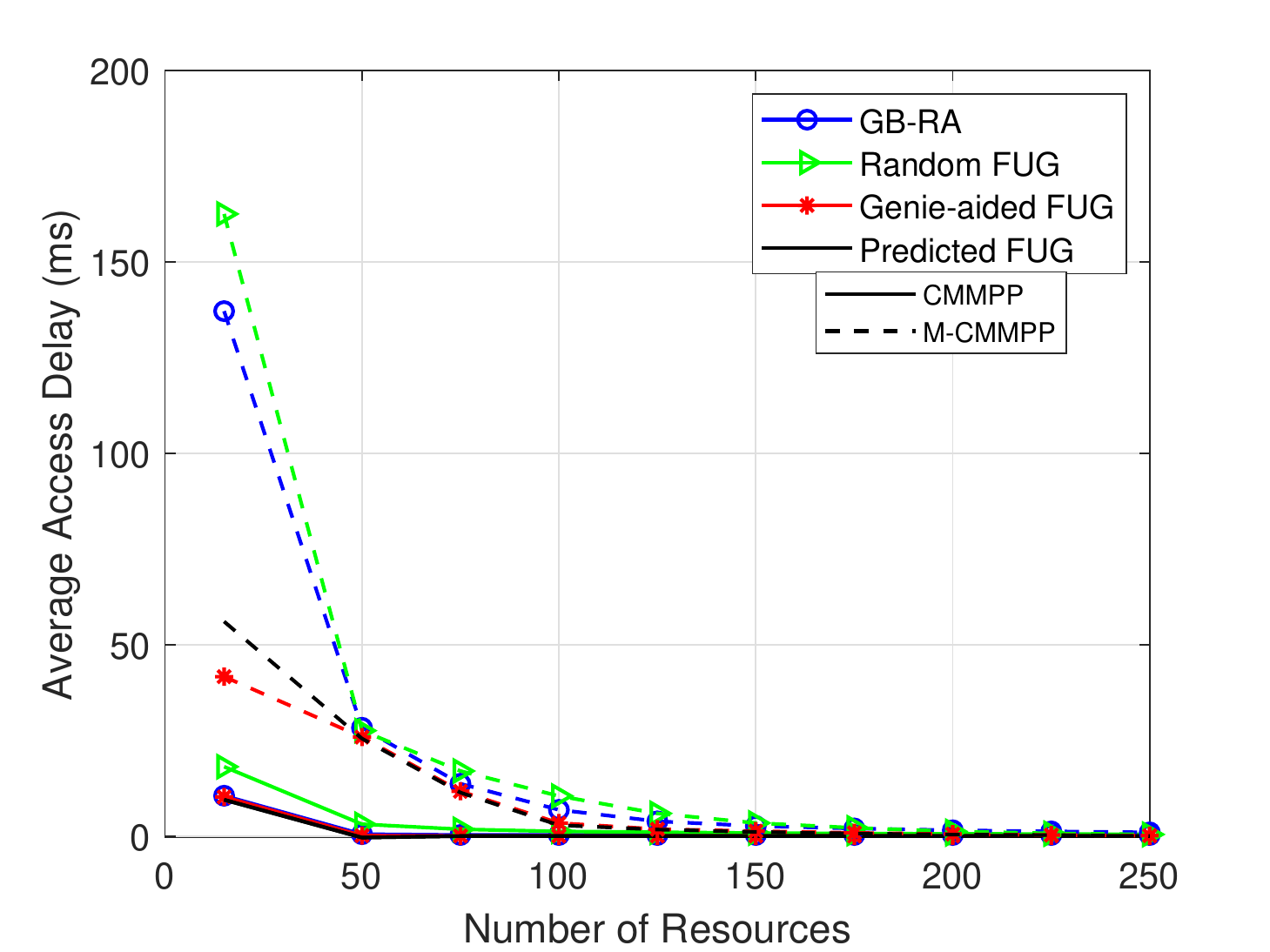}
\vspace{-0mm}
\centering
\caption{Average access delay while adjusting the number of available resources: 1000 MTDs, $\alpha$ = 0.1, $a$ = 0.1, and $T_m = 20 \: ms$. Beyond 75 frequency resources, all schemes converge.}
\label{Total_Access_Delay}
\vspace{-1mm}
\end{figure}

\begin{figure}[!t]
\centering
\includegraphics[width=0.98\columnwidth]{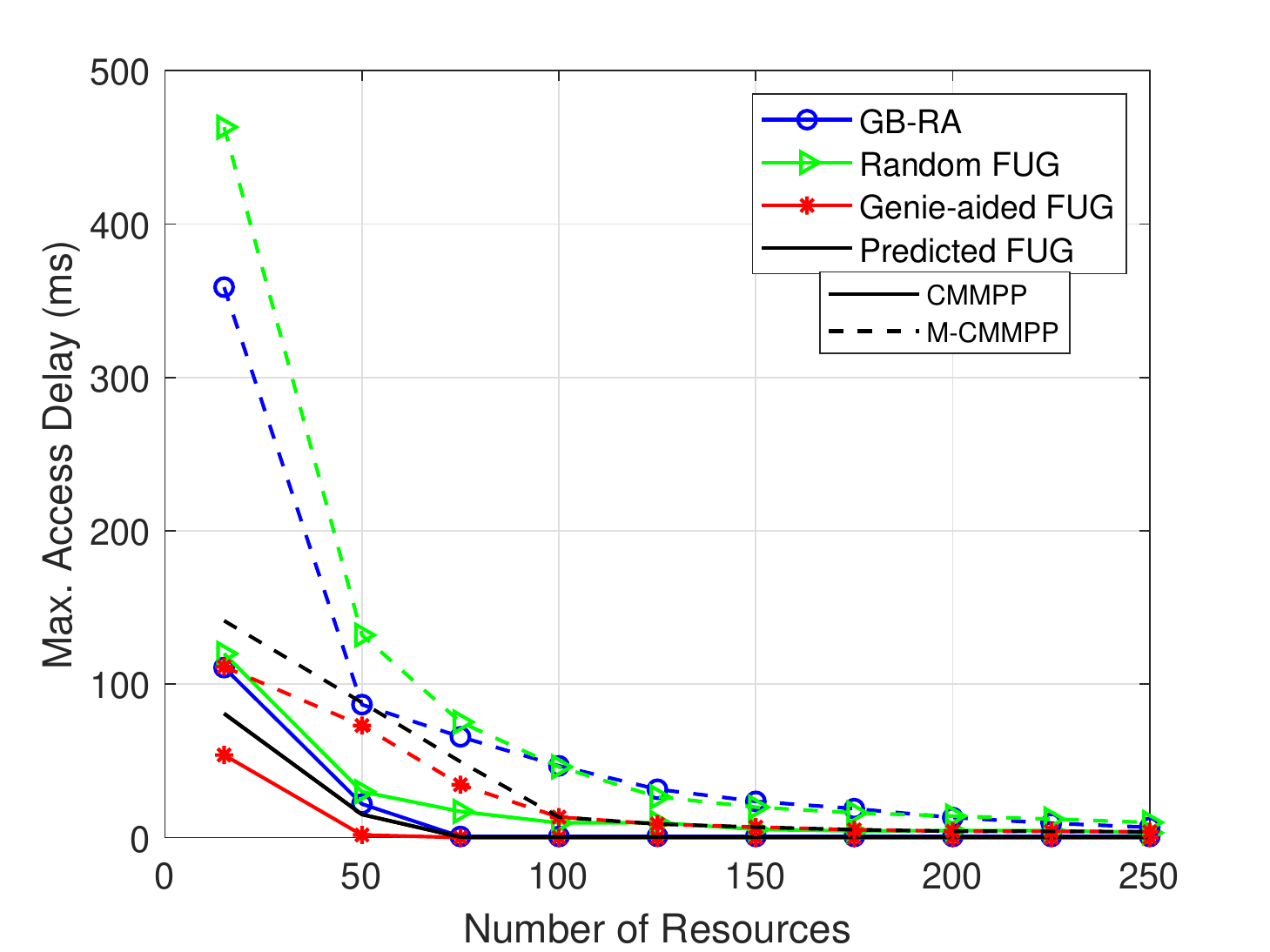}
\vspace{-1mm}
\centering
\caption{Maximum access delay, while adjusting the number of available resources: 1000 MTDs, $\alpha$ = 0.1, $a$ = 0.1, and $T_m = 20 \: ms$. Beyond 75 frequency resources, all schemes converge.}
\label{Max_Delay}
\vspace{-0mm}
\end{figure}

In Figs. \ref{Total_Access_Delay} and \ref{Max_Delay}, the average access delay and maximum access delay are plotted, respectively. The proposed predicted FUG almost approaches the genie-aided FUG. Furthermore, the proposed FUG presents a free-collision resource allocation scheme, whereas GB-RA suffers from several preamble collisions. The random FUG is presented to mention the importance of having a traffic predictor with high accuracy. Having low traffic prediction accuracy can cause even worse performance than the RA schemes. For less than 50 frequency resources, we notice that the predicted FUG algorithm can achieve extremely low latency in the order of 1 to few milliseconds. For this low latency, at least a total of 120 Gbytes of packets are successfully delivered to the BS in the M-CMMPP scenario compared to 108 and 60 Gbytes for the GB-RA and random FUG, respectively as shown in Fig.~\ref{Throughput}. This evidences the outstanding performance of the proposed predictive FUG in the presence of a limited number of resources serving a massive IoT deployment, which is the case of interest.

\vspace{-0.0mm}

\section{Conclusions}\label{conc_sec}
In this paper, we have proposed a novel fast uplink grant resource allocation scheme based on SVM and LSTM. First, we have set up the CMMPP and M-background processes CMMPP models to be our system models. Next, we have implemented an SVM algorithm to classify devices into different priorities. Then, we have developed an LSTM architecture to predict the traffic of a real-time MTD dataset. Afterward, the device classification and traffic prediction were used to schedule the resources. Simulation results have shown that the proposed FUG outperforms RA schemes and almost approaches genie-aided FUG in terms of throughput and latency. We have presented different kinds of error correction techniques that are used to avoid error accumulation. In addition, We have shown the importance of having an accurate traffic predictor to avoid random allocation behavior. Applying learning-based solutions to choose the appropriate exploration rate $\alpha$ is an open research problem for future work.

\vspace{-0mm}
\bibliographystyle{IEEEtran}
\bibliography{di}
\end{document}